\documentclass[journal,draftcls,onecolumn,oneside]{IEEEtran}
\ifCLASSINFOpdf
   \usepackage[pdftex]{graphicx}
   \DeclareGraphicsExtensions{.pdf,.jpeg,.png}
\else
   \usepackage[dvips]{graphicx}
   \DeclareGraphicsExtensions{.eps}
\fi

\usepackage[cmex10]{amsmath}
\usepackage{amssymb}
\usepackage{bm}
\usepackage{mathrsfs}

\usepackage{caption}
\usepackage[T1]{fontenc}

\DeclareMathOperator{\Span}{span}

\DeclareMathOperator{\Vol}{vol}

\newtheorem{Def}{\textbf{Definition}}
\newtheorem{Theorem}{\textbf{Theorem}}
\newtheorem{Lemma}{\textbf{Lemma}}
\newtheorem{Corollary}{\textbf{Corollary}}

\begin{document}
%
\title{Stable Embedding of Grassmann Manifold via Gaussian Random matrices}
%
%
%
\author{Hailong Shi, Hao Zhang, Gang Li, and Xiqin Wang%
\thanks{H. Shi, H. Zhang, G. Li and X. Wang are with the Department of Electronic Engineering, Tsinghua University, Beijing 100084, China.
Emails: shl06@mails.tsinghua.edu.cn, haozhang@mail.tsinghua.edu.cn, gangli@mail.tsinghua.edu.cn, and wangxq\_ee@tsinghua.edu.cn. }%
}%
\maketitle

\begin{abstract}
Compressive Sensing (CS) provides a new perspective for data reduction without compromising performance when the signal of interest is sparse or has intrinsically low-dimensional structure.
The theoretical foundation for most of existing studies on CS is based on the stable embedding (i.e., a distance-preserving property) of vectors that are sparse or in a union of subspaces via random measurement matrices.
To the best of our knowledge, few existing literatures of CS have clearly discussed the stable embedding of linear subspaces via compressive measurement systems.
In this paper, we explore a volume-based stable embedding of multi-dimensional signals based on Grassmann manifold, via Gaussian random measurement matrices.
The Grassmann manifold is a topological space in which each point is a linear vector subspace,
and is widely regarded as an ideal model for multi-dimensional signals generated from linear subspaces.
In this paper,  we formulate the linear subspace spanned by multi-dimensional signal vectors as points on the Grassmann manifold, and use the volume and the product of sines of principal angles (also known as the product of principal sines) as the generalized norm and distance measure for the space of Grassmann manifold.
 We prove a volume-preserving embedding property for points on the Grassmann manifold via Gaussian random measurement matrices, i.e., the volumes of all parallelotopes from a finite set in Grassmann manifold are preserved upon compression. This volume-preserving embedding property is a multi-dimensional generalization of the conventional stable embedding properties, which only concern the approximate preservation of lengths of vectors in certain unions of subspaces.
Additionally, we use the volume-preserving embedding property to explore the stable embedding effect on a generalized distance measure of Grassmann manifold induced from volume.
It is proved that the generalized distance measure, i.e., the product of principal sines between different points on the Grassmann manifold, is well preserved in the compressed domain via Gaussian random measurement matrices.
Numerical simulations are also provided for validation.

\end{abstract}
\begin{IEEEkeywords}
stable embedding, RIP, union of subspaces, Grassmann manifold, principal angle
\end{IEEEkeywords}
\section{Introduction}

\par
Compressive Sensing (CS) \cite{donoho2006compressed}\cite{RobustUncertainty}\cite{eldar2012compressed}\cite{aeron2010information}\cite{baraniuk2010applications} provides a new perspective for data reduction without compromising performance when the signal of interest is sparse or has intrinsically low-dimensional structure.
 Typically the problem of CS is described as $\bm y = \bm \Phi \bm x$,
where $\bm x \in \mathbb{R}^N$ is a k-sparse original signal vector ($\|\bm x\|_0 \leq k, k<<N$), $\bm y \in \mathbb{R}^M( M < N)$ is the compressed measurement vector, and $\bm \Phi \in \mathbb{R}^{M \times N}$ is the measurement matrix (or the sensing matrix).
In the CS literatures, to sufficiently ensure unique signal representation and robust signal recovery, the measurement matrix should approximately preserve the length of all sparse vectors.
i.e., there exists a constant $0< \delta <1$, such that
\begin{equation}\label{RIPIneq}
(1-\delta)\|\bm x\|_2^2 \leq \|\bm \Phi \bm x\|_2^2 \leq (1+\delta)\|\bm x\|_2^2
\end{equation}
holds for all k-sparse vectors $\bm x$ with $\|\bm x\|_0 \leq k$. This expression is the well-known \textit{Restricted Isometry Property} (RIP) of the measurement matrix \cite{Decoding}\cite{SimpleProof}\cite{RIPImplications}.
It can be derived that for two k-sparse vectors $\bm x_1$ and $\bm x_2$ with $\|\bm x_1\|_0 \leq k$ and $\|\bm x_2\|_0 \leq k$, if the measurement matrix $\bm \Phi$ satisfies RIP of order 2k, i.e., (\ref{RIPIneq}) holds for all 2k-sparse vectors, then
\begin{equation}\label{RIPDist}
(1-\delta)\|\bm x_1 - \bm x_2\|^2_2 \leq \|\bm \Phi \bm x_1 - \bm \Phi \bm x_2\|^2_2 \leq (1 + \delta)\|\bm x_1-\bm x_2\|^2_2.
\end{equation}
This means that $\bm \Phi$ approximately preserves the Euclidean distance between any pair of k-sparse vectors. This distance-preserving property in (\ref{RIPDist}) is a more general form of RIP and is commonly referred to as the property of \textit{stable embedding} for sparse vectors \cite{SPComp}.
In addition, there are theoretical results showing that the angles between any pair of sparse vectors
are approximately preserved as well \cite{haupt2007generalized}\cite{Achievable2013Chang}.
\par Furthermore, in \cite{lu2008theory}\cite{gedalyahu2010time}\cite{UnionSubspace}\cite{RobustRecoveryUnion}, the signals of interest in CS has been extended from the conventional sparse vectors to vectors that belong to a union of subspaces.
The unions of subspaces model incorporates many signal models previously considered in original CS settings \cite{UnionSubspace}\cite{RobustRecoveryUnion}, and plays an important role in many subfields of CS, e.g., Multiple Measurement Vector (MMV) in CS \cite{cotter2005sparse}\cite{RobustRecoveryUnion}, Block Sparse Recovery \cite{RobustRecoveryUnion}\cite{duarte2011structured}, and Model-Based Compressive Sensing \cite{baraniuk2010model}.
In \cite{SPComp}\cite{UnionSubspace}\cite{RobustRecoveryUnion}, results analogous to RIP, known as the "$\mathcal A$-RIP" \cite{UnionSubspace} or "Block RIP" \cite{RobustRecoveryUnion}, were proposed. It was proven in \cite{UnionSubspace}\cite{SPComp} that the randomly generated measurement matrix $\bm \Phi$ can approximately preserve the length of a vector as well as the distance between two vectors that lie in a union of subspaces with a notably high probability, i.e., (\ref{RIPIneq}) and (\ref{RIPDist}) hold for all vectors that lie in a union of subspaces.
It is known that this distance-preserving property also ensures the unique signal representation and robust recovery performance of CS for signals from unions of subspaces \cite{UnionSubspace}\cite{RobustRecoveryUnion}, and this property is typically referred to as the stable embedding property for unions of subspaces \cite{SPComp}.
\par
Recently, the stable embedding property was extended to signals modeled as low-dimensional Riemannian sub-manifolds in Euclidean space \cite{newAnalManifold2013}\cite{baraniuk2009random}\cite{StableManifoldEmb2012}. Similar results about the preservation of Euclidean distances of vectors that lie on a low-dimensional sub-manifold via random measurement matrices were proved,  i.e., (\ref{RIPIneq}) and (\ref{RIPDist}) also hold for all vectors that lie on a Riemannian sub-manifold.
In these settings, the Riemannian sub-manifold model is a generalization of the sparse signal model relying on bases or dictionaries \cite{RedundentDict}\cite{candes2011compressed}\cite{elad2010role}\cite{elad2010sparse} and incorporates sophisticated low-dimensional nonlinear geometrical structures.

\par The previous studies on CS mentioned above involve a common stable embedding property of individual vectors, i.e., the preservation of distances (or equivalently lengths) among vectors that are sparse, or lie on a sub-manifold, or belong to a certain union of subspaces, via random measurement matrices. Although the unions of subspaces model is the most popular signal model and is extensively used in various CS applications,
there is few theoretical analysis describing the embedding effect on these linear subspaces via random measurement matrices.
Whereas in this paper, we explore a volume-based stable embedding property to describe the embedding effect on linear subspaces via Gaussian random measurement matrices based on knowledge of Grassmann manifold \cite{absil2004riemannian}.
The Grassmann manifold is a topological space with each point representing a linear subspace, if a linear subspace spanned by multi-dimensional signal vectors is formulated as a point on the Grassmann manifold, a multi-dimensional data matrix will be the basic element representing this point.
The Grassmann manifold is widely regarded as an ideal model for multi-dimensional signals and has been extensively studied in various subfields of signal processing,
 e.g., wireless communication \cite{absil2004riemannian}\cite{inoue2011grassmannian}\cite{love2003grassmannian}\cite{dai2008quantization}\cite{zheng2002communication}, image processing \cite{o2002fitting}\cite{wang2011subspaces}, and machine learning \cite{turaga2008statistical}\cite{hamm2008grassmann}.
The reason why the Grassmann manifold is used to explore the stable embedding of linear subspaces via random measurement matrices is twofold.
First,
the Grassmann manifold has rich topological structure such as geodesics and metrics \cite{absil2004riemannian}, and various distance measures can be defined to describe the relationships between points on the Grassmann manifold \cite{robinson1998separation}\cite{miao1992principal}\cite{qiu2005unitarily}\cite{hamm2008grassmann};
and second, it allows us to formulate and analyze linear subspaces as points in a continuous space,
as a matter of fact, the Grassmann manifold is a natural generalization of the unions of subspaces in the sense that a union of subspaces is actually a subset of several isolated points in the Grassmann manifold.
Thus from this point of view, the Grassmann manifold is intrinsically preferable in our exploration for stable embedding of linear subspaces.

\par It should be mentioned that another important work
by Weiyu Xu and Babak Hassibi \cite{xu2008compressed}\cite{xu2011precise} discussed a certain topic of CS using the Grassmann manifold.
The principal difference between the work of Weiyu Xu et al. in \cite{xu2008compressed}\cite{xu2011precise} and this paper is that,
their analyzes in \cite{xu2008compressed} and \cite{xu2011precise} only involved the conventional vector-form signals, i.e., the approximately sparse signal vectors, and the Grassmann manifold was used as an analytical framework to analyze the null-space property of random measurement matrices \cite{xu2008compressed}; whereas
 our work proves a new volume-based stable embedding property of points on the Grassmann manifold, and reveals a general stable embedding of linear subspaces via Gaussian random measurement matrices.


\par The main contributions of this paper are threefold. First, we formulate multi-dimensional signals as points on the Grassmann manifold, to study the stable embedding of Grassmann manifold via Gaussian random matrices.
This formulation allows us to use volume as a generalized norm function, and the product of principal sines as a generalized distance measure, to describe this general stable embedding of linear subspaces based on Grassmann manifold.
\par
Second, the property of Gaussian random matrices that approximately preserves the volume of all parallelotopes residing in a finite set in Grassmann manifold is proved, and a sufficient condition on the dimension of Gaussian random measurement matrices to guarantee this corresponding stable embedding is given.
To the best of our knowledge, this volume-preserving embedding property has not been discussed previously, and this novelty is one of the main contributions of our work.
The volume is chosen as a generalized metric or distance measure of points on the Grassmann manifold, in order to explore the stable embedding of linear subspaces via Gaussian random measurement matrices.
The reason for the choice of volume is that,
in conventional Euclidean space, each point is a vector and the metric measure is induced by the vector norm function, whereas for a linear subspace, a set of linearly independent vectors spanning this subspace, i.e., a basis, is commonly used to specify this subspace; therefore, we can treat the volume of the parallelotope spanned by a set of vectors as a multi-dimensional generalization of the norm (or length) of an individual vector.
Volume is a key characteristic for the space of Grassmann manifold.
Typically, the volume of parallelotopes spanned by the bases of subspaces has been used to provide a measure of separation between different subspaces \cite{robinson1998separation}\cite{miao1992principal}; and as we know that  principal angles provide a wide class of metrics and distance measures on the Grassmann manifold \cite{qiu2005unitarily}, the volume is also closely related to the principal angles between subspaces \cite{miao1992principal}.
Motivated by these factors, we use the volume as a generalized norm function of points on the Grassmann manifold, and prove the volume-preserving embedding property of Grassmann manifold.
This volume-based stable embedding property, analogous to the RIP and stable embedding property based on length, is given in a probabilistic formulation, i.e., this volume-preserving property is satisfied with a notably high probability under a certain condition on the dimension of measurement matrices.
We provide a rigorous proof of this volume-based stable embedding property, as well as discussions on its differences from and connections with the previous result of RIP \cite{SimpleProof} and stable embedding of unions of subspaces \cite{UnionSubspace}\cite{SPComp}.
To derive our result, we use such techniques as the theory of random matrices to derive the concentration inequality for the determinant of random matrices, and knowledge of high-dimensional geometry to obtain an improved result of covering numbers, as well as the matrix perturbation theory and the union bound.
It is shown that the result is a high-dimensional generalization of the results of stable embedding for unions of subspaces and RIP. Indeed, if we only consider 1-dimensional "parallelotopes" in our theorem, the volume-preserving embedding property reduces back to the conventional length-preserving embedding property for individual vectors lying in certain unions of subspaces.
\par Third, using the theorem of volume-based stable embedding, we also derive a theorem to describe the stable embedding effect on a generalized distance measure, i.e., the product of principal sines, between points on the Grassmann manifold, via Gaussian random measurement matrices.
It is shown that our generalized distance measure, i.e., the product of principal sines, can be directly derived from volume. Then we prove that the product of principal sines is theoretically preserved via Gaussian random measurement matrices using knowledge of our volume-based stable embedding property.

\par Throughout this paper, we use small bold letters $\bm x$ to denote vectors, capital bold letters $\bm X$ to denote matrices; we use $\|\bm X\|_p$ and $\|\bm x\|_p$
to denote the $\ell_p$ norm of the matrix $\bm X$ and vector $\bm x$, and use $\bm I_d$ to denote
the identity matrix of dimension $d$. $\Span(\bm X)$ is used for representation of the linear subspace spanned by column vectors of the matrix $\bm X$, and $[\bm X,\bm Y]$ for the juxtaposition of the matrix $\bm X$ and $\bm Y$. $\mathbb{P}$ and $\mathbb{E}$ denotes the probability and expectation respectively.

\par
The remainder of this paper is organized as follows.
First, in Section II, necessary definitions, such as the Grassmann manifold, volume, principal angles, and stable embedding based on length of vectors are presented.
Next, the main results of this paper, i.e., the theorem for the volume-based stable embedding property of Grassmann manifold, as well as the stable embedding effect on a generalized distance measure for points on the Grassmann manifold, is stated and discussed in Section III.
The sketched proof of our main results is provided in section IV, and finally detailed proofs are included in appendices.

\section{Preliminary Background}
\subsection{Grassmann Manifold and Unions of Subspaces}

\par The unions of linear subspaces model is a general signal model commonly used in CS \cite{RobustRecoveryUnion}\cite{cotter2005sparse}\cite{duarte2011structured}\cite{baraniuk2010model}. The signal $\bm x$ in this model is assumed as a vector from a union of linear subspaces, defined as \cite{lu2008theory}\cite{UnionSubspace}
\begin{equation}\label{UoS}
\mathcal{X}=\bigcup_{i=1}^L \mathcal{X}_i \subset \mathbb{R}^N,\quad \mathcal{X}_i = \{\bm x = \bm X_i \bm \alpha_i , \bm X_i \in \mathbb{R}^{N\times k},\bm \alpha_i \in \mathbb{R}^{k}\},
\end{equation}
where the matrix $\bm X_i$'s column vectors form the basis of the corresponding subspace $\mathcal{X}_i$, with $\Span(\bm X _i)=\mathcal{X}_i$, and $ \dim(\mathcal{X}_i) = k < N$.
The unions of linear subspaces model is a generalization of the conventional sparse model (for the sparse model, the columns of $\bm X_i$'s are the canonical bases and $L = \binom{N}{k}$) and incorporates many signal models in the conventional Compressive Sensing settings.
\par
The Grassmann manifold $\text{Gr}(k,N)$ is defined as a topological space in which each point is a $k$-dimensional linear vector subspace of $\mathbb{R}^N$(or $\mathbb{C}^N$). In general, a union of subspaces in (\ref{UoS}) is equivalently a finite collection of different points in $\text{Gr}(k,N)$, that is,
\begin{equation}\label{GrassSets}
\boldsymbol{\mathcal{G}}(k,N,L) := \{\mathcal{X}_1,\cdots,\mathcal{X}_L\},\quad \mathcal{X}_i \in \text{Gr}(k,N),1\leq i \leq L.
\end{equation}
As far as we know, although the unions of subspaces model is quite general and offers extensive applications in various fields of CS, there is no theoretical analysis describing the relationships between these subspaces and the implication of their relationships in CS, whereas the Grassmann manifold enables us to describe these relationships by exploiting its topological structure.
As in \cite{hamm2008grassmann}\cite{robinson1998separation}\cite{miao1992principal}\cite{qiu2005unitarily}, different metrics and distance measures have been used to describe the topological structure of the Grassmann manifold. From this point of view, the Grassmann manifold is intrinsically preferable for describing relationships between subspaces, and enables the study on stable embedding of subspaces.


\subsection{Stable Embedding Property for Unions of Subspaces}
The stable embedding of unions of subspaces, also equivalently referred to as "$\mathcal{A}$-RIP"\cite{UnionSubspace} or "Block-RIP"\cite{RobustRecoveryUnion}, describes the length-preserving embedding property of vectors in a certain union of subspaces via compressive measurement matrices \cite{SPComp}\cite{UnionSubspace}. A well-known sufficient condition for the stable embedding property via Gaussian random measurement matrices was given by M.E Davies et al. in 2009 \cite{UnionSubspace} and stated that, for i.i.d. Gaussian random matrices $\bm \Phi \in \mathbb{R}^{M \times N}$ with each entry $\phi_{i,j}$ satisfies
\begin{equation}\label{MomentCond}
\phi_{i,j} \thicksim \mathcal{N}(0,\frac{1}{M}),
\end{equation}
if for any $t>0$, and any constant $0 < \delta <1$,
\begin{equation}\label{StableEmbBnd}
M \geq \frac{2}{c\delta}\Big(\log(2L) +k \log\Big(\frac{12}{\delta} \Big) + t \Big),
\end{equation}
then the property of length-preservation
\begin{equation}\label{StableEmbedding}
(1-\delta)\|\bm x \|_2^2 \leq \|\bm \Phi\bm x \|_2^2 \leq (1+\delta)\|\bm x \|_2^2,
\end{equation}
holds for all vectors in a union of subspaces $\bm x \in \mathcal{X}=\bigcup_{i}^L \mathcal{X}_i$ with probability
\begin{equation}
\mathbb{P} \geq 1 - e^{-t}.
\end{equation}
\par
As is known, this length-preserving embedding property of vectors in unions of subspaces via Gaussian random sensing matrices can be equivalently generalized to the distance-preserving embedding property in \cite{UnionSubspace}: \par
For i.i.d. Gaussian random matrices $\bm \Phi \in \mathbb{R}^{M \times N}$ with each entry satisfying (\ref{MomentCond}), for any $t>0$, and any constant $0 < \delta <1$, let
\begin{equation}\label{StableEmbBnd}
M \geq \frac{2}{c\delta}\Big(\log(2\bar L) +k \log\Big(\frac{12}{\delta} \Big) + t \Big),
\end{equation}
where $\bar L = L(L-1)/2$, then the property of distance-preservation
\begin{equation}\label{StableEmbedding}
(1-\delta)\|\bm x_1 - \bm x_2\|_2^2 \leq \|\bm \Phi\bm x_1 - \bm \Phi\bm x_2 \|_2^2 \leq (1+\delta)\|\bm x_1 -\bm x_2 \|_2^2,
\end{equation}
holds for all vectors $\bm x_1 , \bm x_2$ in a union of subspaces with probability
$\mathbb{P} \geq 1 - e^{-t}$.

\par
\subsection{Volumes in the Grassmann manifold}
As is known, any element of $\text{Gr}(k,N)$, i.e., any $k$-dimensional linear subspace $\mathcal{X} \subset \mathbb{R}^N$ is usually specified by a matrix of full column rank
\begin{equation}
\bm X = [\bm x_1, \bm x_2, \cdots, \bm x_k] \in \mathbb{R}^{N \times k}, k<N,
\end{equation}
with columns forming the basis of the corresponding subspace, i.e., $\Span(\bm X) = \mathcal{X} \in \text{Gr}(k,N)$.
\par The $d$-dimensional volume of a full-rank matrix $\bm S = [\bm s_1,\cdots \bm s_d] \in \mathbb{R}^{N \times d}$, with $1 \leq d \leq k$ and $\Span(\bm S) \subset \mathcal{X} \in \text{Gr}(k,N)$, is defined as \cite{ben1992volume}
\begin{equation}\label{VolumeDef1}
\Vol_d (\bm S):= \prod_{i=1}^d \sigma_i,
\end{equation}
where $\sigma_1 \geq \sigma_2 \geq \cdots \geq \sigma_d > 0$ are singular values of matrix $\bm S$.  The volume of the matrix $\bm S$ is also referred to as the $d$-dimensional parallelotope spanned by the column vectors of $\bm S$. Because $\bm S$ is of full column rank, the volume is equivalently \cite{ben1992volume}\cite{miao1992principal}
\begin{equation}\label{VolumeDef2}
\Vol_d (\bm S) = \sqrt{\det(\bm{S^TS})}.
\end{equation}

\par Particularly, if $d=1$, $\bm S = [\bm s_1]$, $\Vol_d( \bm S)$ equals $\|\bm s_1\|_2$, i.e., the length of this single vector; if $d=2$, $\Vol_d( \bm S)$ becomes the area of the parallelogram spanned by the two vectors $\bm s_1$ and $\bm s_2$, and if $d=3$, $\Vol_d( \bm S)$ is the volume of the parallelepiped spanned by the three vectors $\bm s_1, \bm s_2$, and $\bm s_3$. From this point of view, we can say that the volume of a parallelotope is a multi-dimensional generalization of the length of a vector. For convenience, we call $\Vol_d( \bm S)$ in (\ref{VolumeDef1}) the volume of subspace $\Span(\bm S)$ corresponding to matrix $\bm S \in \mathbb{R}^{N \times d}$.
\par Volume is an important quantity in the Grassmann manifold space, it provides a measure of separation between two linear subspaces and is closely related to the principal angles between subspaces \cite{miao1992principal}\cite{miao1996product}. In fact, for any two $k$-dimensional linear subspaces $\mathcal{X},\mathcal{Y}$ with $\mathcal{X}\bigcap \mathcal{Y} = \{0\}$
 and spanned by columns of matrices $\bm X$ and $\bm Y$, the principal angles $\pi/2 \geq \theta_1,\cdots \geq \theta_k > 0$ between $\mathcal{X}$ and $\mathcal{Y}$ satisfy\cite{miao1992principal}
\begin{equation}\label{volumeangle}
\Vol_{2k}([\bm X, \bm Y]) = \Vol_k(\bm X)\Vol_k(\bm Y)\cdot \prod_{i=1}^k \sin \theta_i,
\end{equation}
where we refer to the expression $\prod_{i=1}^k \sin \theta_i$ as \textit{the product of principal sines} \cite{miao1992principal}.
Indeed, we can define a wide class of metric measures using the principal angles\cite{qiu2005unitarily}\cite{hamm2008grassmann}, e.g., the geodesic distance
$
d_G(\mathcal{X},\mathcal{Y})=\sum_{i=1}^k \theta_i^2,
$
and the projection distance
$
d_P(\mathcal{X},\mathcal{Y}) = \big( \sum_{i=1}^k \sin^2 \theta_i \big)^{1/2}.
$
 According to \cite{qiu2005unitarily}, various measure functions that may not be as strict as metrics (which must satisfy the triangle inequality) also can be used as distance measures for different points on the Grassmann manifold, and following the terminology used in \cite{qiu2005unitarily}, without verifying the triangle inequality, we choose the product of principal sines induced by the volume in (\ref{volumeangle}) as a generalized distance measure on the Grassmann manifold in the following analyzes.

\section{Main Results}
\subsection{Formulating Multi-dimensional Signals as Points on the Grassmann manifold}
\par The definition of Grassmann manifold indicates that it is preferable to study multi-dimensional signals generated from linear subspaces. In this section, we introduce the formulation of multi-dimensional signals as points on the Grassmann manifold.
This formulation implies that, the basic element to be received and processed will be a multi-dimensional data matrix, with columns containing an array of different sampled vectors, and the definition in terms of \textit{signals on the Grassmann manifold} will be:
\begin{Def}
The multi-dimensional data matrix received from the signal acquisition front-end
\begin{equation}\label{setofvector}
\bm X = [\bm x_1, \cdots,\bm x_k] \in \mathbb{R}^{N \times k},\quad \bm x_i \in \mathbb{R}^N,1\leq i \leq k,
\end{equation}
is called \textit{a signal on the Grassmann manifold}, where $\bm x_i$'s are different sampled vectors composing this multi-dimensional signal.
\end{Def}
\par Generally, these $\bm x_i$ are linearly independent, thus we have $\Span(\bm X) \in \text{Gr}(k,N)$, and each data matrix $\bm X$ will specify a point on the Grassmann manifold $\text{Gr}(k,N)$; therefore a signal on the Grassmann manifold is represented by the data matrix $\bm X$ as in (\ref{setofvector}).\par
A simple example of this formulation can be found in \cite{zheng2002communication}. In the multiple-antenna communication systems, there exist $M$ transmit and $N$ receive antennas with $M \leq N$, and the channel fading coefficients form a $N \times M$ matrix $\bm H$, the received multi-dimensional signal over a period of $D$ ($D > M$) samples from the $N$ receive antennas can be written in a matrix form:
$$\bm Y = \bm{HX}+\bm W,$$
 where $\bm X \in \mathbb{R}^{M \times D}$, with row vectors $\bm x_i \in \mathbb{R}^D$ corresponding to the transmitted data at the $i$th transmit antenna and $\bm Y \in \mathbb{R}^{N \times D}$ with rows $\bm y_j \in \mathbb{R}^D$ corresponding to the received data for the $j$th received antenna.  In addition, $\bm W \in \mathbb{R}^{M \times D}$ denotes the additive noise.
The data matrix $\bm X^T$ can be formulated as a signal on the Grassmann manifold $\text{Gr}(M,D)$ and $\bm Y^T$ as the version of $\bm X^T$ corrupted by noise $\bm W$. This is a typical example of the formulation of signals on the Grassmann manifold.
\par For another famous example in \cite{hamm2008grassmann}, in the subspace-based learning problems, where the data to be learned and classified are generated from linear subspaces, data matrices as in (\ref{setofvector}) are formulated as signals on the Grassmann manifold. Then various metric functions in Grassmann manifold can be used as kernel functions, to enhance the learning and classifying performance of Linear Discriminant Analysis \cite{hamm2008grassmann}.
\par Similar to (\ref{setofvector}), we also formulate multi-dimensional signals in the compressed domain in terms of \textit{compressed measurement signals on the Grassmann manifold},
and what is received as an element from the compressive measurement front-end is also a multi-dimensional data matrix, the definition is:
\begin{Def}
The data matrix from the compressive measurement front-end formed as
\begin{equation}\label{Compdata}
\bm Y =[\bm y_1, \cdots, \bm y_k]= [\bm \Phi\bm x_1, \cdots,\bm \Phi \bm x_k] \in \mathbb{R}^{M \times k},
\end{equation}
is called \textit{a compressed measurement signal on the Grassmann manifold}, where $\bm \Phi \in \mathbb{R}^{M \times N},M<N$ is the measurement matrix, and $\bm x_j$'s are different orignal signal vectors before compression.
\end{Def}
\par As is mentioned, in most general settings of CS, the original signal vectors are supposed to lie in a union of subspaces, i.e., a finite set in Grassmann manifold.
Thus the original signal on Grassmann manifold specifies a point $\mathcal{X}_i(1\leq i \leq L)$ in a finite set $\boldsymbol{\mathcal{G}}(k,N,L)$ as in (\ref{GrassSets}),
and the compressed measurement signal on the Grassmann manifold, i.e., $\bm Y$, specifies a point $\bm \Phi\mathcal{X}_i$ $(1\leq i \leq L)$ in another finite set
\begin{equation}
\boldsymbol{\mathcal{G}}'(k,M,L):= \{\bm \Phi\mathcal{X}_1,\cdots,\bm \Phi\mathcal{X}_L\},
\end{equation}
where
$\bm \Phi\mathcal{X}_i:= \Span(\bm \Phi \bm X_i)\subset \text{Gr}(k,M)$ \footnote{It is noted that for the random matrix $\bm \Phi$, if $k$ is sufficiently small, the dimension of the subspace $\bm \Phi\mathcal{X}_i$ is the same as $\mathcal{X}_i$ almost surely. So it will be a general assumption throughout this paper that $\dim(\bm \Phi\mathcal{X}_i)=\dim(\mathcal{X}_i)=k$.} represents the subspaces transformed by the measurement matrix $\bm \Phi$.
\par
Our objective in this paper is to study the stable embedding with respect to these two finite sets on the Grassmann manifold, i.e., the set of signals on the Grassmann manifold $\boldsymbol{\mathcal{G}}(k,N,L) $ and the set of compressed measurement signals on the Grassmann manifold $\boldsymbol{\mathcal{G}}'(k,M,L)$. \par
Next, we will use the volume in (\ref{VolumeDef1}) as a generalized norm function, and the product of principal sines in (\ref{volumeangle}) as a generalized distance measure, to explore the stable embedding of points in a finite set in Grassmann manifolds.
Before we start, a definition of the general stable embedding property of Grassmann Manifold based on volumes is required:

\begin{Def}(volume-based stable embedding property)\label{DefSVE}
We say that the measurement matrix $\bm \Phi$ provides a volume-based stable embedding of a finite set in Grassmann manifold, i.e., $\boldsymbol{\mathcal{G}}(k,N,L) \subset \text{Gr}(k,N)$, with the dimension of volume $d$ ($d \leq k$) and coefficient $(A,\varepsilon)$, if for every matrix $\bm S\in\mathbb{R}^{N\times{d}}$  with $\Span(\bm S)\subset \mathcal{X}_i\in\boldsymbol{\mathcal{G}}(k,N,L),1\leq i \leq L$, we have
\begin{equation}
\left|\log \frac{\Vol_d(\bm \Phi \bm S)}{\Vol_d(\bm S)}-A\right|\leq\varepsilon,
\end{equation}
alternately,
\begin{equation}
A-\varepsilon\leq\log(\Vol_d(\bm \Phi \bm S))-\log(\Vol_d(\bm S))\leq{A+\varepsilon}.
\end{equation}
\end{Def}
We will show that this definition of volume-based stable embedding property will be supported by theoretical results from the following several theorems.
\subsection{The Volume-based Stable Grassmann Manifold Embedding}
\begin{Theorem}\label{MainThm}
Consider a finite set in Grassmann manifold $\boldsymbol{\mathcal{G}}(k,N,L)$ and a random matrix $\bm \Phi \in \mathbb{R}^{M \times N}$ with elements $\phi_{i,j}$ being i.i.d Gaussian random variables with mean 0 and variance $1/M$; for any constant $0 < C_s < 1$ and any integer $1\leq d \leq k$, for the matrix
$$
\bm S = [\bm s_1,\cdots,\bm s_d] \in \mathbb{R}^{N \times d},\ \|\bm s_j\|_2=1,\ 1 \leq j \leq d,
$$
where
$$
\Span(\bm S)\subset\mathcal{X}_i\in\boldsymbol{\mathcal{G}}(k,N,L), \ 1 \leq i \leq L,
$$
satisfying $\Vol_d(\bm S)> C_s$, we have
\begin{equation}\label{VolConcentrationExp}
\mathbb{E} \log \frac{\Vol_d(\bm \Phi \bm S)}{\Vol_d(\bm S)}  =  \frac{1}{2} \sum_{p=1}^{d}\Big(\psi[(M-p+1)/2]+\log 2 - \log M\Big).
\end{equation}
And there exist constants $\delta_s>0$, and $C,C'>0$, only depending on $C_s$, such that for any $0 < \varepsilon<d^\frac{3}{2}\delta_s(1+C')$ and $t>0$, if
\begin{align}\label{MeasBound}
M\geq\frac{4(1+C')^2(1+C) \cdot d}{\varepsilon^2} \Big[ \log(2L) + d(\frac{3}{2}k-1) \log(e \cdot d)
+ d\cdot k \log(\lceil\frac{3(1+C')}{\varepsilon}\rceil) +t\Big]+d-1,
\end{align}
then
\begin{equation}\label{VolConcentrationMain}
\left|\log \frac{\Vol_d(\bm \Phi \bm S)}{\Vol_d(\bm S)}-\mathbb{E}\log\frac{\Vol_d(\bm \Phi\bm S)}{\Vol_d(\bm S)}\right| \leq \varepsilon
\end{equation}
holds for every matrix $\bm S$ with probability
\begin{equation}
\mathbb{P} \geq 1 - e^{-t},
\end{equation}
where $\psi(x) = \frac{\partial}{\partial z}\log \Gamma(z)|_{z=x}$ is the Digamma function (for Digamma function, refer to \cite{TCai2013arXiv}).
\end{Theorem}
\par Theorem \ref{MainThm} describes the approximately volume-preserving property of a finite set in Grassmann manifold via Gaussian random measurement matrices.  A sufficient condition on $M$, i.e., the number of compressive measurements, in (\ref{MeasBound}) to guarantee the volume preservation in (\ref{VolConcentrationMain}) is given in Theorem \ref{MainThm}.  If $M$ is bounded as (\ref{MeasBound}), the volumes of all matrices from the finite set in  Grassmann manifold can be approximately preserved with an overwhelming probability, as in (\ref{VolConcentrationMain}). Here are some further discussion:
\par 1)
The matrices discussed in Theorem \ref{MainThm} are conditioned to have unit-norm column vectors, i.e., $\|\bm s_j\|_2=1,1 \leq j \leq d$.
This constraint is for convenience of proof and implies no loss of generality; actually, if there is any column $\bm s_j$ of $\bm S = [\bm s_1,\cdots,\bm s_d]$ that is not unit-norm,
such as $\|\bm s_j\|_2=c \neq 1$, then the volume of the column-normalized matrix $\bm{\hat S} = [\bm s_1,\cdots,\bm s_j/\|\bm s_j\|_2,\cdots,\bm s_d]$ will be $\Vol_d{\bm{\hat S}} = c^{-1} \cdot \Vol_d(\bm S)$, the only difference is a multiplication of a constant. Therefore, it is sufficient that we only consider the parallelotopes spanned by unit-norm vectors.
\par 2)
An axillary parameter $C_s$ is introduced in Theorem \ref{MainThm}. It is the lower bound of the volume of matrix $\bm S \in \mathbb{R}^{N \times d}$ to ensure the validity of conclusion.
Indeed, for fixed $\varepsilon>0$, if $C_s$ becomes smaller, then $C'$ will become larger, causing the lower bound in (\ref{MeasBound}) to increase, meaning that the stable embedding is more difficult to achieve for smaller volumes.
In fact, if the volume of $\bm S$ is too small, i.e., $\Vol_d(\bm S)$ is tending to zero, then the dimension of the corresponding subspace $\Span(\bm S)$ will become less than $d$. The volume-preserving properties for dimension $d$ are somewhat meaningless for these subspaces with dimension less than $d$.

\par 3)
The main result of volume preservation is shown in (\ref{VolConcentrationMain}) and (\ref{VolConcentrationExp}).
The parameter $A$ and $\varepsilon$ from Definition \ref{DefSVE} can be easily derived from (\ref{VolConcentrationMain}).
Furthermore, if $M$ satisfies the bound in (\ref{MeasBound}), then log ratio of $\Vol_d(\bm \Phi \bm S)$ and $\Vol_d(\bm S)$ will concentrate around its expectation
\begin{equation}\label{Concenter}
\frac{1}{2} \sum_{p=1}^{d}\Big(\psi[(M-p+1)/2]+\log 2 - \log M\Big),
\end{equation}

It should be noted that this expectation value depends only on $M$ and $d$, so $N>M$ is not relevant here.
\begin{figure}[htbp]
\centering
\includegraphics[width=0.9\textwidth]{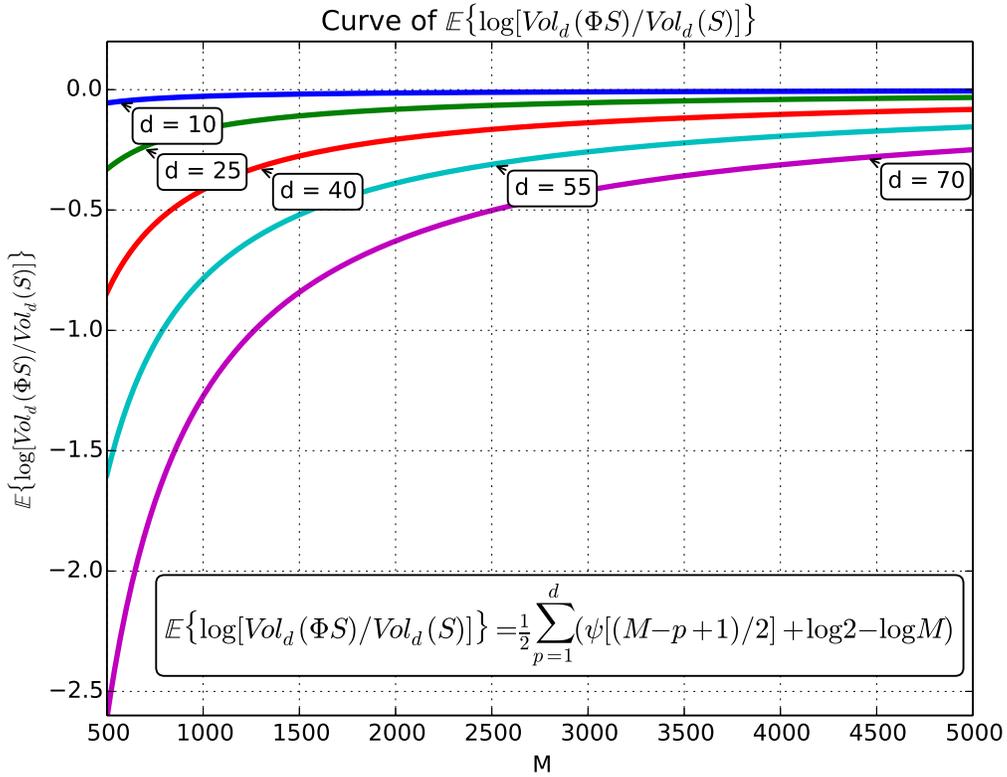}
\caption{the expectation curve the log ratio of volumes $\Vol_d(\bm \Phi \bm S)$ and $\Vol_d(\bm S)$ in which we choose $M$ from 500 to 5000 and $d$ from 10 to 70}
		\label{figure1}
\end{figure}
The curve of (\ref{Concenter}) is plotted in Figure \ref{figure1}, where $M$ ranges from 500 to 5000 and $d$ ranges from 10 to 70. It can be observed that the value of (\ref{Concenter}) is slightly less than 0, which means the effects of the random measurement matrix on the volume
of subspaces are slightly "biased". and by "biased", we mean the log ratio of $\Vol_d(\bm \Phi \bm S)$ and $\Vol_d(\bm S)$ does not concentrate approximately around 0 but around (\ref{Concenter}).
Additionally, as $M$ increases, (\ref{Concenter}) grows closer to 0, which indicates that more measurements produces less "bias" of the volumes of subspace. However, when $d$ becomes larger, (\ref{Concenter}) deviates away from 0, which means the volume preservation becomes worse when the dimension of subspace increase.
\par
Indeed, if we use asymptotic expansion \cite{TCai2013arXiv} of the Digamma function $\psi(x)$, which is $\psi(x) = \log x - \frac{1}{2x}+O(\frac{1}{|x|^2})$, then we have
\begin{equation}\label{taylorexp}
\frac{1}{2} \sum_{p=1}^{d}\Big(\psi[(M-p+1)/2]+\log 2 - \log M\Big) = \frac{1}{2} \sum_{p=1}^{d}\Big(\log(M-p+1) - \log M - \frac{1}{M-p+1}  + O(\frac{1}{(M-p+1)^2})\Big),
\end{equation}
and it can be observed that as $M \rightarrow \infty$ and $d/M < \infty$, (\ref{Concenter}) will tend to 0, and as $d$ grows larger, (\ref{Concenter}) will tend away from 0. This explains the curve in Figure \ref{figure1}.
\par 4)
As is shown, Theorem \ref{MainThm} describes the volume-preserving embedding for all matrices with a given number of columns $d$, different values of $d$ determines different measurement bounds in (\ref{MeasBound}) as well as different concentration inequalities in (\ref{VolConcentrationMain}).
Particularly, if $d=1$, the 1-dimensional volume is length, i.e., $\Vol_1(\bm s) = \|\bm s\|_2$, and we obtain
\begin{equation}\label{dimension1exp}
\mathbb{E} \log \frac{\|\bm \Phi \bm s\|_2}{\|\bm s\|_2} = \frac{1}{2} \Big(\psi[M/2]+\log 2 - \log M\Big),
\end{equation}
and if
\begin{equation}\label{dimension1bd}
M \geq \frac{4(1+C')^2(1+C)}{\varepsilon^2}\Big[\log(2L) + (\frac{3}{2}k-1)\log(e) + k \log(\lceil\frac{3(1+C')}{\varepsilon}\rceil) +t\Big],
\end{equation}
then
\begin{equation}\label{dimension1res}
\left| \log \frac{\|\bm \Phi \bm s\|_2}{\|\bm s\|_2}  - \mathbb{E}\big\{ \log \frac{\|\bm \Phi \bm s\|_2}{\|\bm s\|_2} \big\}\right| \leq \varepsilon
\end{equation}
holds with probability of at least $1-e^{-t}$.
\par Compared with the length-preserving embedding of unions of subspaces proposed by Davies et al, the measurement bound in (\ref{dimension1bd})  shows a little difference with (\ref{StableEmbBnd}).
The main reason for these differences is that we use a different approximation method to analyze the probabilistic concentration of volumes of multi-dimensional parallelotopes, and this method may be slightly rougher for the 1-dimensional "parallelotope". As a whole, the measurement bound (\ref{dimension1bd}) for $d=1$ is of the same order with (\ref{StableEmbBnd}) by Davies et al.
\par
In addition, it appears in (\ref{dimension1exp}) that $\psi[M/2]+\log 2 - \log M$ is less than 0, which means
\begin{equation}\label{explogratio}
\mathbb{E}(\log \frac{\|\bm \Phi \bm s\|_2^2}{\|\bm s\|_2^2}) < 0,
\end{equation}
and the result by Davies and Baraniuk et al. states \cite{UnionSubspace}\cite{SimpleProof}
\begin{equation}\label{expratio}
\mathbb{E}(\frac{\|\bm \Phi \bm s\|_2^2}{\|\bm s\|_2^2}) = 1.
\end{equation}
The reason is that what we focus on is the concentration of the log ratio of $\|\bm \Phi \bm s\|_2$ and $\|\bm s\|_2$, and the difference between (\ref{explogratio}) and (\ref{expratio}) can be explained by Jensen's Inequality, i.e.,
\begin{equation}\label{jensen}
 \psi[M/2]+\log 2 - \log M = \mathbb{E}(\log \frac{\|\bm \Phi \bm s\|_2^2}{\|\bm s\|_2^2}) \leq \log  \mathbb{E}( \frac{\|\bm \Phi \bm s\|_2^2}{\|\bm s\|_2^2}) = 0.
\end{equation}
 In brief, the result of Theorem \ref{MainThm} for 1-dimensional "parallelotopes" reduces back to the length-preserving embedding of unions of subspaces proposed by Davies et al, whereas Theorem \ref{MainThm} can be further extended to multi-dimensional scenarios.

\par 5)
The bound in (\ref{MeasBound}) is the sufficient condition for a Gaussian random matrix $\bm \Phi \in \mathbb{R}^{M \times N}$ to provide the volume-preserving embedding property. Here $M$ should be of the order of:
\begin{equation}
M \thicksim O(d \cdot \log(L) + d^2 \cdot k\log (e\cdot d)).
\end{equation}
Particularly, when $d=1$,
\begin{equation}
M \thicksim O(\log(L) +k ),
\end{equation}
which coincides with the result of stable embedding for unions of subspaces by Davies et al. Additionally, if $d=k$, then $M $ should be of the order of:
\begin{equation}
M \thicksim O(k \cdot \log(L) + k^3\log(k)).
\end{equation}
These results indicate that we require additional compressive measurements to ensure the volume-based stable embedding property. \par
To be more specific, if we consider the conventional sparse model, if $L = \binom{N}{k} \leq (eN/k)^k$, then $M $ should be of the order of:
\begin{equation}\label{RIPm}
M \thicksim O(d \cdot k \cdot \log(N/k) + d^2 \cdot k\log(e\cdot d)),
\end{equation}
and if $d=1$, (\ref{RIPm}) becomes the conventional RIP result, i.e., $M \thicksim O(k \cdot \log(N/k))$.

\subsection{Effect of stable embedding on a generalized distance measure for Grassmann manifold}
In this section, we discuss the effect of the volume-preserving embedding on a generalized distance measure of compressed measurement signals on the Grassmann manifold.
 Without loss of generality, we prefer to consider each point in the original set in Grassmann manifold to be disjoint, which means different points in $\boldsymbol{\mathcal{G}}(k,N,L) = \{\mathcal{X}_1,\cdots,\mathcal{X}_L\}$ satisfy $\mathcal{X}_i \bigcap \mathcal{X}_j = \{0\}, i \neq j$
\footnote{If $\mathcal{X}_i \bigcap \mathcal{X}_j \neq \{0\}$, different methods exists to address the relationships between principal angles and volumes. These relationships are slightly complicated and trivial, so we simply focus on the most typical $\mathcal{X}_i \bigcap \mathcal{X}_j = \{0\}$ scenario and leave the $\mathcal{X}_i \bigcap \mathcal{X}_j \neq \{0\}$ for future work.} \footnote{The result in Theorem \ref{MainThm} as well as the result in Corollary \ref{Corr1} will ensure that $\bm \Phi\mathcal{X}_i \bigcap \bm \Phi\mathcal{X}_j = \{0\}$.}.
Before we present the second theorem, a corollary, which is derived from Theorem \ref{MainThm}, is presented first.
\begin{Corollary}\label{Corr1}
Consider the $\bar L: = L(L-1)/2$ pairs of subspaces $\mathcal{X}_i \oplus \mathcal{X}_j$ from the finite set in Grassmann manifold $\boldsymbol{\mathcal{G}}(k,N,L)$, with $\mathcal{X}_i \bigcap \mathcal{X}_j = \{0\}, i \neq j$, and a random matrix $\bm \Phi \in \mathbb{R}^{M \times N}$ with elements $\phi_{i,j}$ being i.i.d Gaussian random variables with mean 0 and variance $1/M$; for any constant $0 < C_s < 1$ , and for every matrix
$$\bm X = [\bm x_1,\cdots,\bm x_d]\in \mathbb{R}^{N\times d},\ \|\bm x_l\|_2=1, 1 \leq l \leq d$$
 with number of columns $1 \leq d \leq 2k$, where $\Span(\bm X) \subset \mathcal{X}_i \oplus \mathcal{X}_j$ and satisfying $\Vol_{d}(\bm X)>C_s$, we have
\begin{eqnarray}
\mathbb{E}\big\{ \log \frac{\Vol_{d}(\bm \Phi \bm X)}{\Vol_{d}(\bm X)} \big\} &=&  \frac{1}{2} \sum_{p=1}^{d}\Big(\psi[(M-p+1)/2]+\log 2 - \log M\Big), \label{expfor2k}
\end{eqnarray}
and there exists $\delta_s>0$, and $C,C'>0$, only depending on $C_s$, such that for any $0 < \varepsilon<\delta_s(1+C')$, if:
\begin{equation}
M \geq \frac{8(1+C')^2(1+C)\cdot k}{\varepsilon^2}\Big[\log(2\bar{L}) + 2k\cdot(3k-1) \log(2e k) + 4k^2 \log(\lceil\frac{3(1+C')}{\varepsilon}\rceil) +\log(2k)+t\Big]+2k-1,
\end{equation}
then
\begin{eqnarray}
\left| \log \frac{\Vol_{d}(\bm \Phi \bm X)}{\Vol_{d}(\bm X)}   - \mathbb{E} \log \frac{\Vol_{d}(\bm \Phi \bm X)}{\Vol_{d}(\bm X)} \right| \leq \varepsilon,\label{conctr2k}
\end{eqnarray}
holds with probability
\begin{equation}
\mathbb{P} \geq 1 - e^{-t},
\end{equation}
where $\psi(x)$ is the Digamma function.
\end{Corollary}
\par This corollary states a similar probabilistic result of the volume-preserving embedding property for all dimensions $1\leq d \leq 2k$, instead of the result for any given dimension in Theorem \ref{MainThm}. According to Corollary \ref{Corr1}, we obtain the second main result of this paper:
\begin{Theorem}\label{MainPrinAgl}
Consider the $\bar L: = L(L-1)/2$ pairs of subspaces $\mathcal{X}_i \oplus \mathcal{X}_j$ from the finite set in Grassmann manifold $\boldsymbol{\mathcal{G}}(k,N,L)$, with $\mathcal{X}_i \bigcap \mathcal{X}_j = \{0\}, i \neq j$, and a measurement matrix $\bm \Phi \in \mathbb{R}^{M \times N}$ which satisfies the volume-preserving embedding property for all dimensions $1\leq d \leq 2k$, i.e., $\bm \Phi$ satisfies corollary \ref{Corr1};
then the principal angles denoted by $\pi/2 \geq\theta_1(\mathcal{X}_i, \mathcal{X}_j) \geq \cdots \geq \theta_k(\mathcal{X}_i, \mathcal{X}_j)> 0$ between $\mathcal{X}_i$ and $\mathcal{X}_j$, as well as the principal angles $\pi/2 \geq\theta_1(\bm \Phi \mathcal{X}_i, \bm \Phi\mathcal{X}_j) \geq \cdots \geq \theta_k(\bm \Phi\mathcal{X}_i, \bm \Phi\mathcal{X}_j) > 0$ between $\bm \Phi \mathcal{X}_i$ and $\bm \Phi \mathcal{X}_j$ for $1\leq i \neq j \leq L$ will satisfy:
\begin{equation}\label{princpAgl}
\left| \log \frac{\prod_m^k\sin\theta_m(\bm \Phi\mathcal{X}_i, \bm \Phi\mathcal{X}_j)}{\prod_m^k\sin\theta_m(\mathcal{X}_i, \mathcal{X}_j)} - \frac{1}{2}\sum_{p=1}^{k}\Big(\psi[(M-p-k+1)/2]-\psi[(M-p+1)/2]\Big)\right| \leq 3\varepsilon,
\end{equation}
where $\psi(x)$is the Digamma function.
\end{Theorem}
\par Theorem \ref{MainPrinAgl} describes the effect of the volume-preserving embedding in Theorem \ref{MainThm} on the generalized distance measure of Grassmann manifold.
It is proved that, the product of principal sines between points on the Grassmann manifold is theoretically approximately preserved, as is shown in (\ref{princpAgl}).
Similar to previous results, the log ratio of $\prod_m^k\sin\theta_m(\bm \Phi\mathcal{X}_i, \bm \Phi\mathcal{X}_j)$ and $\prod_m^k\sin\theta_m(\mathcal{X}_i, \mathcal{X}_j)$ in (\ref{princpAgl}) concentrates around a center, which is
\begin{equation}\label{AglCtr}
\frac{1}{2}\sum_{p=1}^{k}\Big(\psi[(M-p-k+1)/2]-\psi[(M-p+1)/2]\Big).
\end{equation}
It also appears that (\ref{AglCtr}) is slightly less than 0, and if $M \rightarrow \infty$ and $k/M < \infty$, (\ref{AglCtr}) will tend to 0.
\par
\begin{figure}
\centering
    \begin{minipage}[htbp]{0.9\textwidth}
    \centering
    \includegraphics*[width=0.9\textwidth]{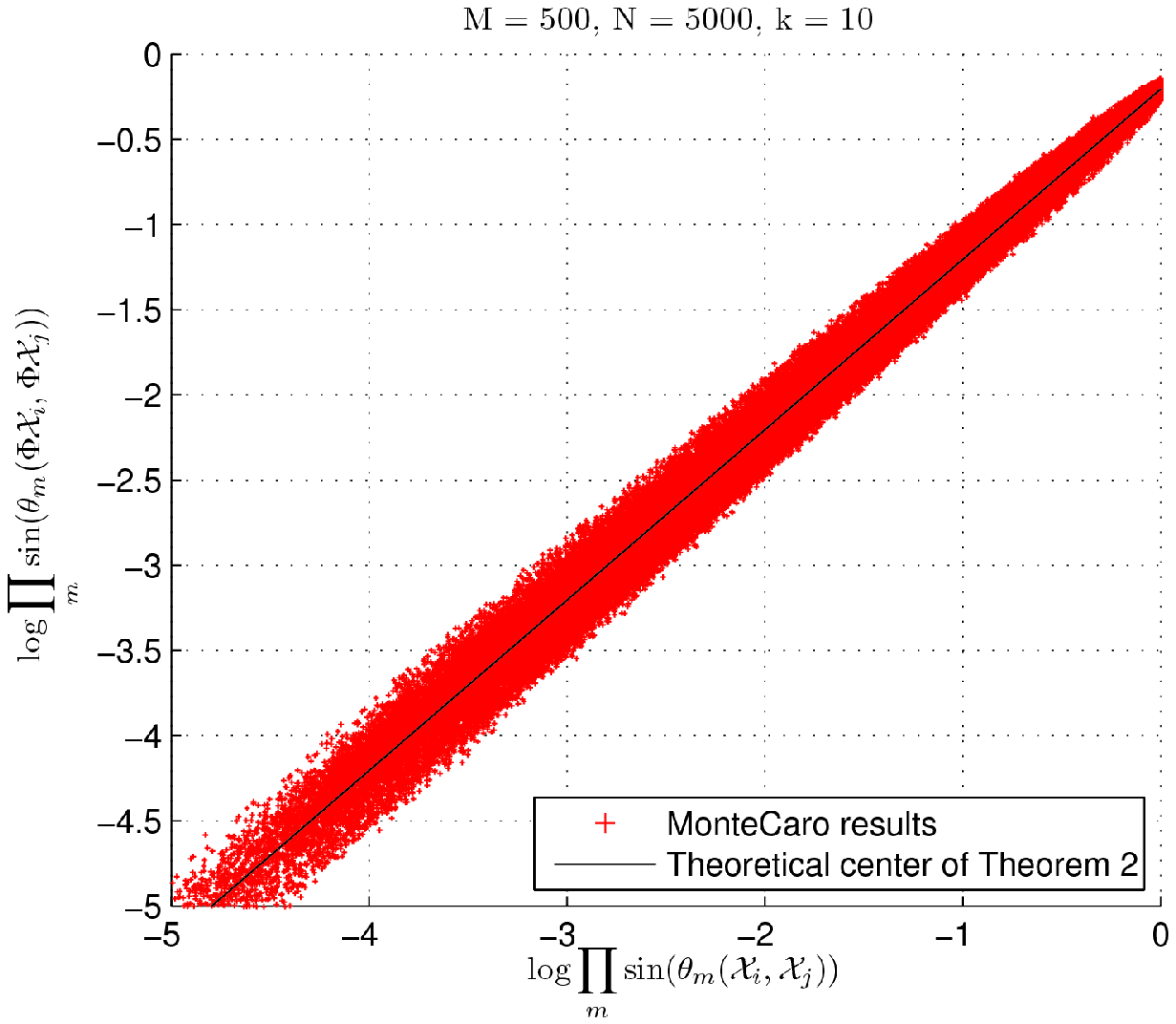}
    \caption{Monte-Carlo simulation result for $\prod_m^k\sin\theta_m(\mathcal{X}_i, \mathcal{X}_j)$ and
 $\prod_m^k\sin\theta_m(\bm \Phi\mathcal{X}_i, \bm \Phi\mathcal{X}_j)$ as well as the theoretical center in (\ref{princpAgl}), and $M=500,k=10$}
		\label{figure7}
    \end{minipage}
    \begin{minipage}[htbp]{0.9\textwidth}
    \centering
    \includegraphics*[width=0.9\textwidth]{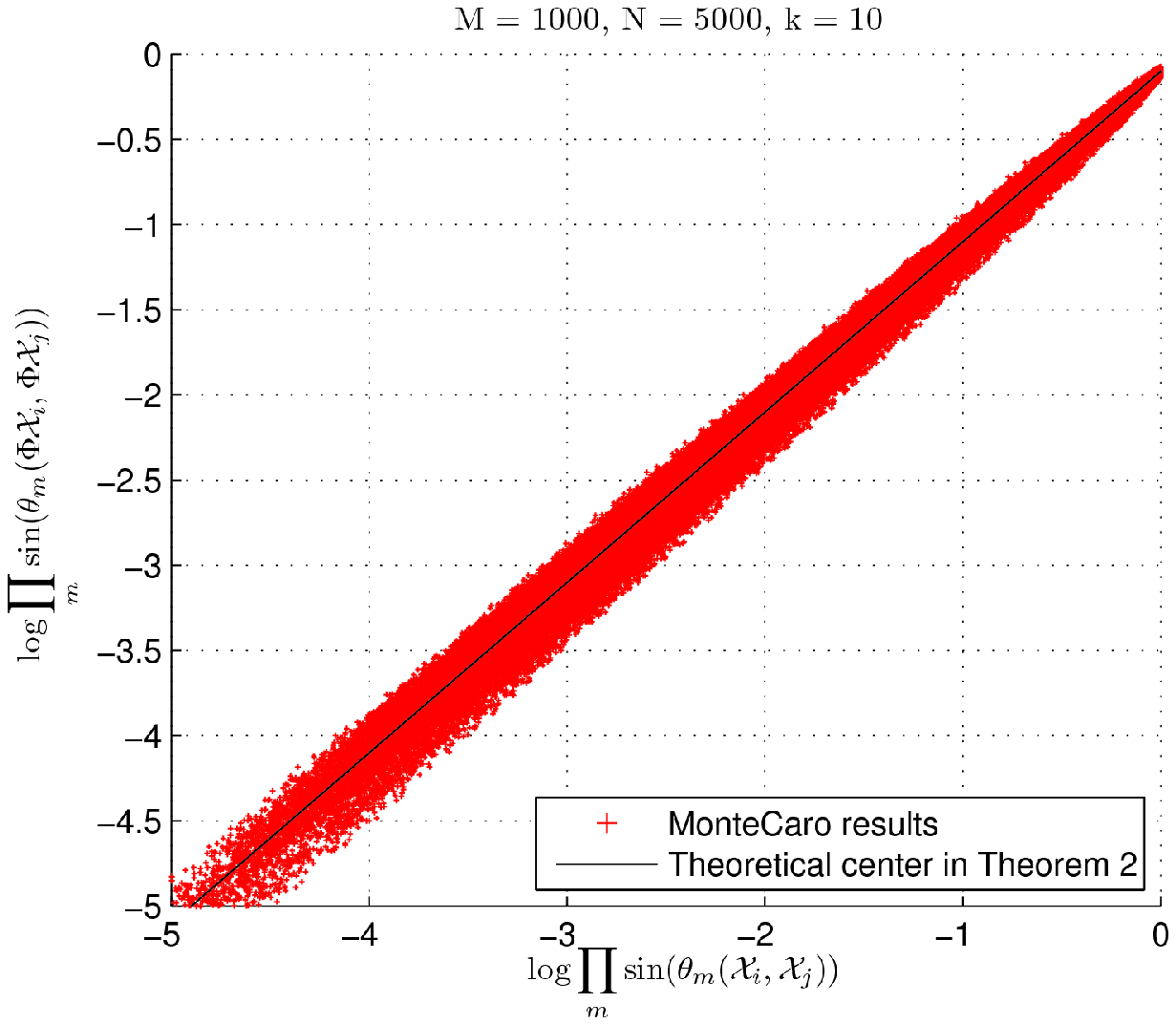}
    \caption{Monte-Carlo simulation result for $\prod_m^k\sin\theta_m(\mathcal{X}_i, \mathcal{X}_j)$ and
 $\prod_m^k\sin\theta_m(\bm \Phi\mathcal{X}_i, \bm \Phi\mathcal{X}_j)$ as well as the theoretical center in (\ref{princpAgl}), and $M=1000,k=10$}
		\label{figure8}
    \end{minipage}
\end{figure}
\begin{figure}
\centering
    \begin{minipage}[htbp]{0.9\textwidth}
    \centering
    \includegraphics*[width=0.9\textwidth]{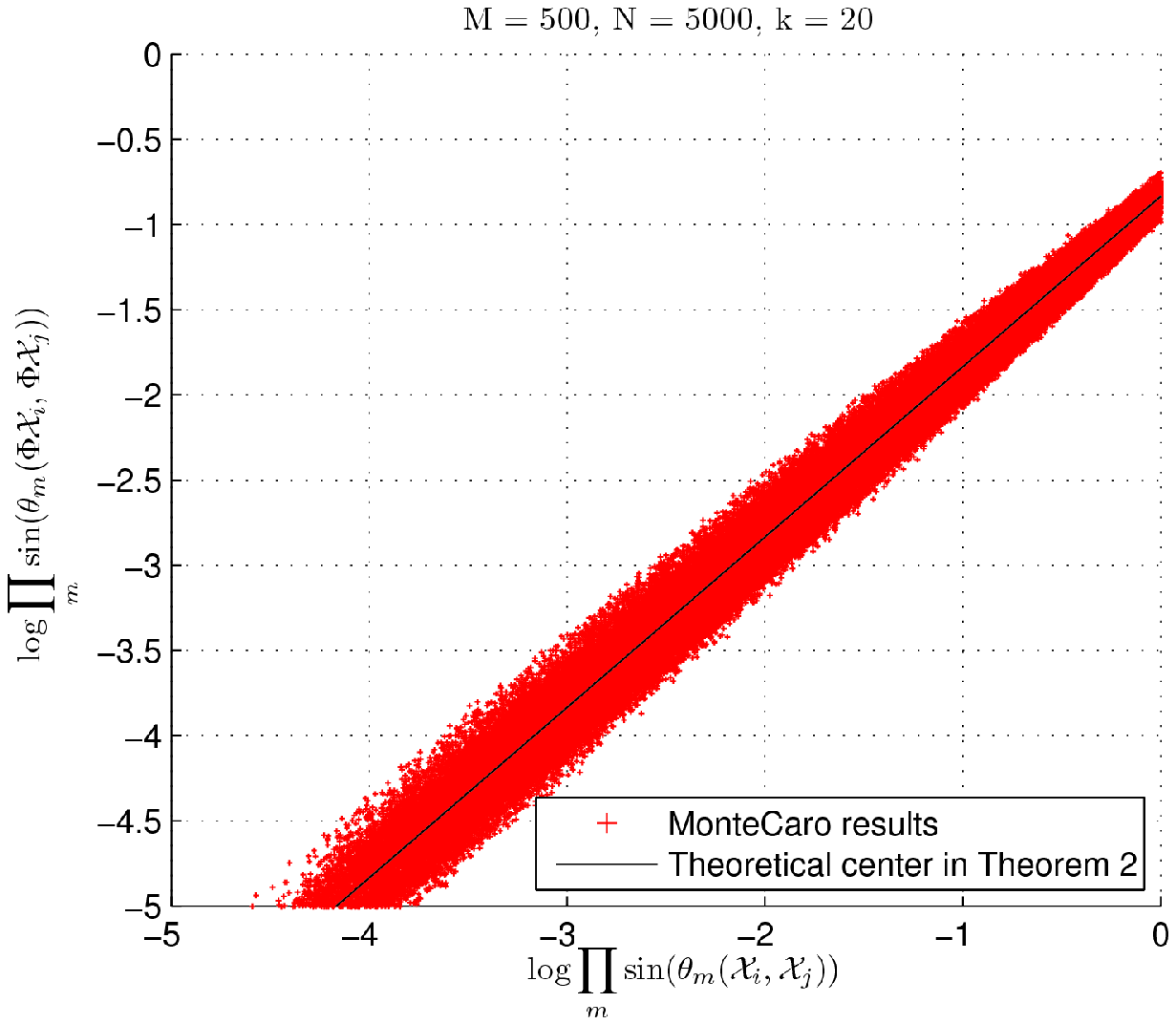}
    \caption{Monte-Carlo simulation result for $\prod_m^k\sin\theta_m(\mathcal{X}_i, \mathcal{X}_j)$ and
 $\prod_m^k\sin\theta_m(\bm \Phi\mathcal{X}_i, \bm \Phi\mathcal{X}_j)$ as well as the theoretical center in (\ref{princpAgl}), and $M=500,k=20$}
		\label{figure9}
    \end{minipage}
    \begin{minipage}[htbp]{0.9\textwidth}
    \centering
    \includegraphics*[width=0.9\textwidth]{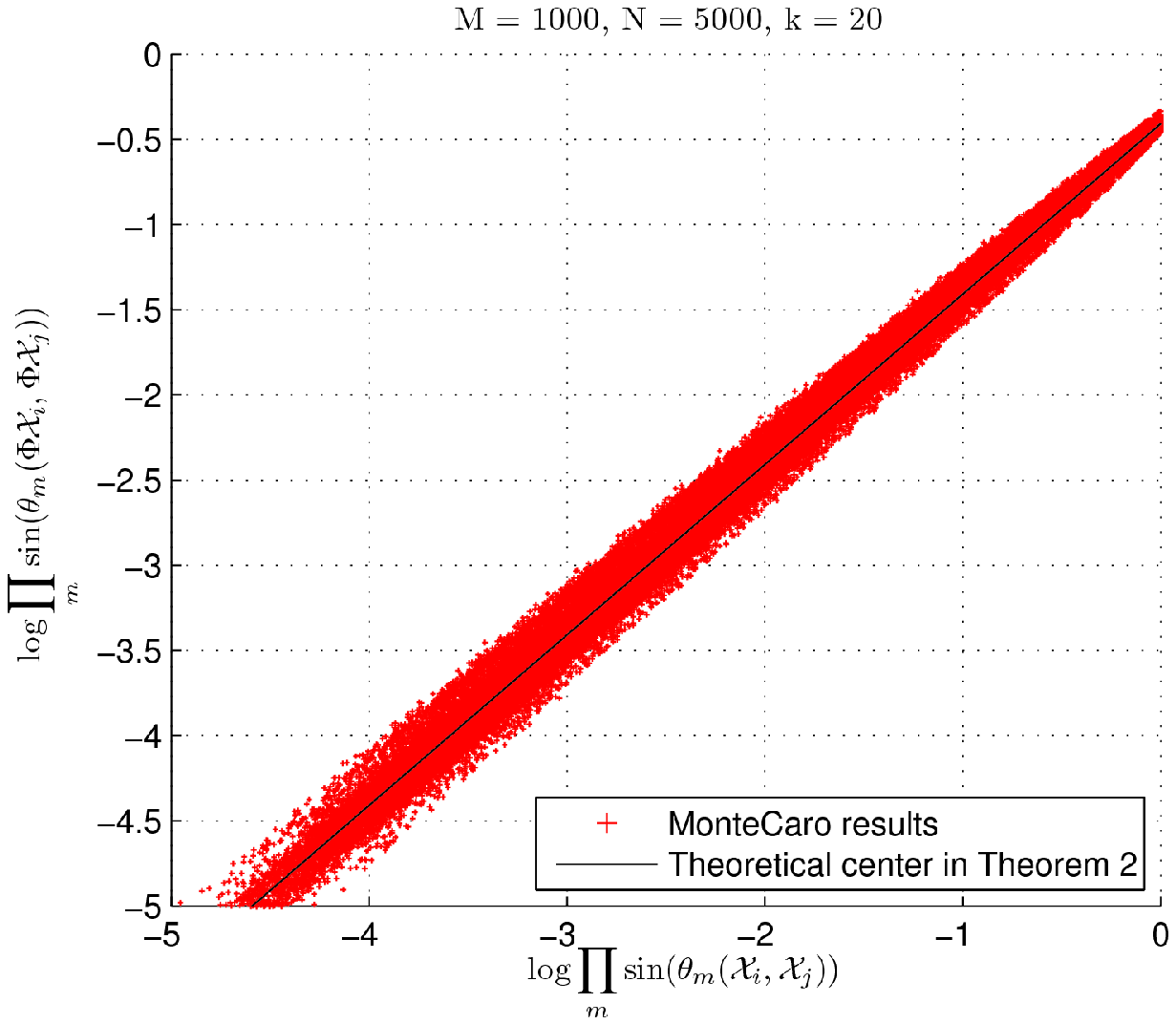}
    \caption{Monte-Carlo simulation result for $\prod_m^k\sin\theta_m(\mathcal{X}_i, \mathcal{X}_j)$ and
 $\prod_m^k\sin\theta_m(\bm \Phi\mathcal{X}_i, \bm \Phi\mathcal{X}_j)$ as well as the theoretical center in (\ref{princpAgl}), and $M=1000,k=20$}
		\label{figure10}
    \end{minipage}
\end{figure}

The Monte-Carlo simulation results verifying the result of Theorem \ref{MainPrinAgl}
are demonstrated in Figure \ref{figure7} to Figure \ref{figure10}, inspired by the simulation strategy in \cite{Achievable2013Chang}. In the simulation, we choose
a randomly generated measurement matrix $\bm \Phi \in \mathbb{R}^{M \times N}$,
with each entry $\phi_{ij}$ independently drawn from $\mathcal{N}(0,1/M)$; and
typically, we choose $N=5000$, and the number of measurements $M$ as $500$ and $1000$. For each $\bm \Phi$, we generate $800$ sets of randomly chosen principal angles $\theta_1,\cdots \theta_k$ under the constraint $\log\prod_m^k\sin\theta_m(\mathcal{X}_i, \mathcal{X}_j)\geq-5$.
And for each set of angles, 100 arbitrary pairs of points $\mathcal{X}_i$ and $\mathcal{X}_j$ on $\text{Gr}(k,N)$ are generated,
with dimensions $k$ equal to $10$ and $20$, respectively.
For each test pair $\mathcal{X}_i$ and $\mathcal{X}_j$, the values of
$\log\prod_m^k\sin\theta_m(\mathcal{X}_i, \mathcal{X}_j)$ and
 $\log\prod_m^k\sin\theta_m(\bm \Phi\mathcal{X}_i, \bm \Phi\mathcal{X}_j)$
 as well as the theoretical center (\ref{AglCtr}) are plotted in these figures. From these figures we can clearly verify
 the result of Theorem \ref{MainPrinAgl}.
\par It can be observed that from Theorem \ref{MainPrinAgl}, we obtain a theoretical guarantee for the close relationship between $\prod_m^k\sin\theta_m(\bm \Phi\mathcal{X}_i, \bm \Phi\mathcal{X}_j)$ and $\prod_m^k\sin\theta_m(\mathcal{X}_i, \mathcal{X}_j)$. Because we know that
\begin{eqnarray}
\prod_m^k\sin\theta_m(\bm \Phi\mathcal{X}_i, \bm \Phi\mathcal{X}_j) &=& \frac{\Vol_{2k}(\bm \Phi[\bm X_i, \bm X_j])}{\Vol_{k}(\bm \Phi \bm X_i)\Vol_{k}(\bm \Phi \bm X_j)},\label{volumerel} \\
\prod_m^k\sin\theta_m(\mathcal{X}_i, \mathcal{X}_j) &=& \frac{\Vol_{2k}([\bm X_i, \bm X_j])}{\Vol_{k}(\bm X_i)\Vol_{k}(\bm X_j)},
\end{eqnarray}
therefore,  we can use (\ref{volumerel}) to measure the distance between different compressed measurement signals on the Grassmann manifold specified by the data matrices $\bm Y_i = \bm \Phi \bm X_i$ and $\bm Y_j = \bm \Phi \bm X_j$. Using this distance measure as in (\ref{volumerel})
has intrinsic advantages. First, it is easy to calculate,
we only need to calculate a determinant directly on the received data matrix $\bm Y_i$, $\bm Y_j$ and $[\bm Y_i, \bm Y_j]$.
Second, as mentioned, the relationship of this distance measure for $\boldsymbol{\mathcal{G}}'(k,M,L)$ with the distance measure for original $\boldsymbol{\mathcal{G}}(k,N,L)$ is theoretically preserved by Theorem \ref{MainPrinAgl}. Thus, we believe that the distance measure in (\ref{volumerel}) is both theoretically trustworthy and computationally efficient.
\section{Proof of the main theorem}
\subsection{Proof of Theorem \ref{MainThm}}
\begin{figure}[htbp]
\centering
\includegraphics*[width=0.9\textwidth]{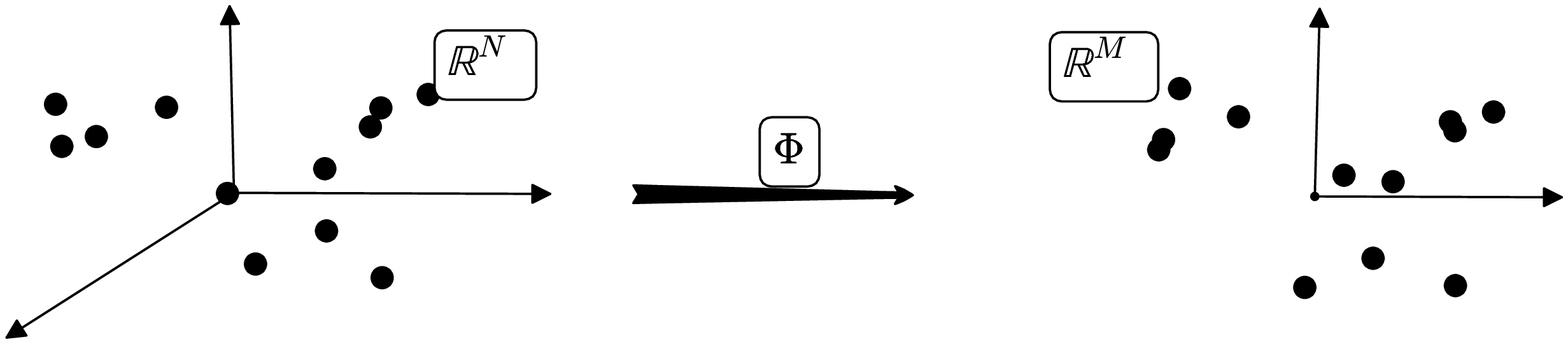}
\caption{The mutual distance between Euclidean points is approximately preserved by the measurement matrix $\bm \Phi$ via the stable embedding property}
		\label{figure2}
\end{figure}

This section presents the proof of Theorem \ref{MainThm}. Motivated by (\ref{StableEmbedding}) and (\ref{StableEmbBnd}) proposed by Davies et al. in \cite{UnionSubspace}, we know that the Gaussian random measurement matrix $\bm \Phi$ can approximately preserve the distances between all pairs of vectors in union of subspaces with tremendous high probability.
This intuitively implies that the volume of subspace spanned by these mutually distance-preserved vectors also should be approximately preserved, as demonstrated in Figure \ref{figure2}.
This is just the statement of Theorem \ref{MainThm}.
\par Our proof of Theorem \ref{MainThm} includes three steps, namely, the concentration inequality, the covering number, and the union bound. In each step, several lemmas will be given as intermediate conclusions.
\subsubsection{Step 1. The Concentration Inequality}
The main conclusion of this step is:
\begin{Lemma}\label{StepConcent}
For any full rank matrix $\bm S\in \mathbb{R}^{N \times d},N > d$ and random matrix $\bm \Phi \in \mathbb{R}^{M \times N}$ with elements $\phi_{i,j}$ being i.i.d Gaussian random variables with mean 0 and variance $1/M$;  the volumes $\Vol_d (\bm S)$ and $\Vol_d (\bm \Phi \bm S)$ will satisfy
\begin{align}\label{detconcent}
&\mathbb{P}\left\{\left|\log \frac{\Vol_d (\bm \Phi \bm S)}{\Vol_d (\bm S)}-\mathbb{E} \log \frac{\Vol_d(\bm \Phi \bm S)}{\Vol_d(\bm S)}\right| \leq \varepsilon \right\} \nonumber\\
\geq& 1 - 2 \exp \left\{-\varepsilon^2/\left(4\sum_{p=1}^d[\frac{1}{M-p+1}+ C\frac{1}{(M-p+1)^2}]\right)\right\}
\end{align}
holds for any $\varepsilon > 0$, where $C>0$ is a constant parameter,
$$
\mathbb{E}\big\{ \log \frac{\Vol_d(\bm \Phi \bm S)}{\Vol_d(\bm S)} \big\} = \frac{1}{2} \sum_{p=1}^d \Big(\psi[(M-p+1)/2]+\log 2 - \log M\Big),
$$
and $\psi(x) = \frac{\partial}{\partial z}\log \Gamma(z)|_{z=x}$ is the Digamma function.
\end{Lemma}
\text{Proof of Lemma \ref{StepConcent}: See Appendix \ref{App1}.}
\par
\begin{figure}[htbp]
\centering
\includegraphics*[width=0.9\textwidth]{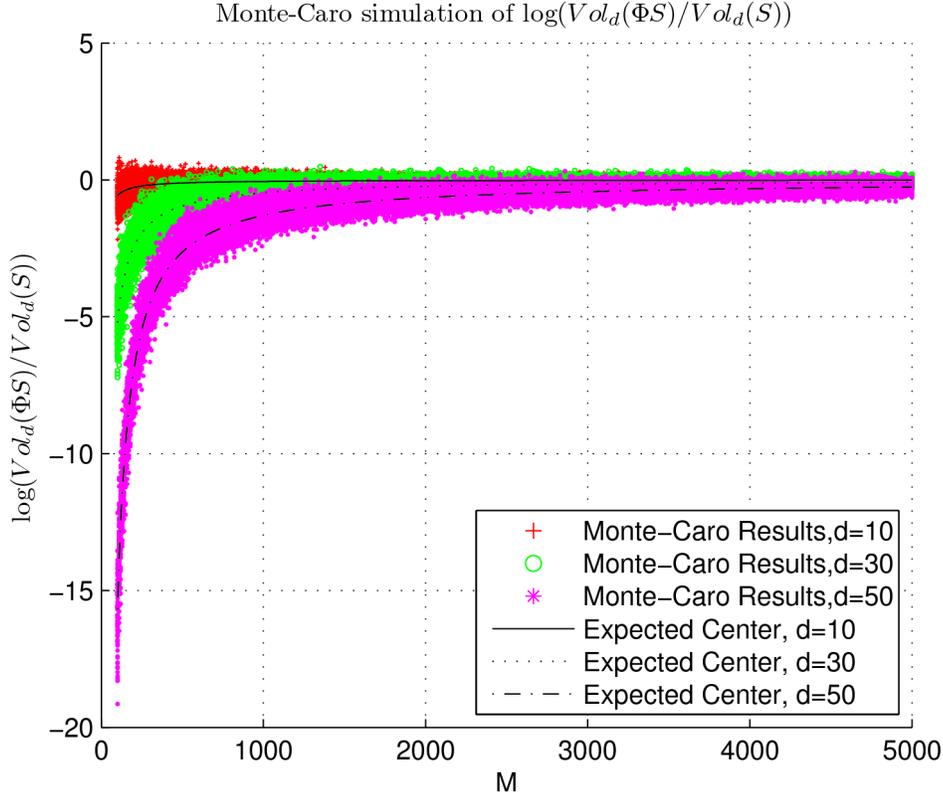}
\caption{Monte-Carlo simulations for the distribution of values of $\log(\Vol_d (\bm \Phi \bm S)/\Vol_d (\bm S))$, where $\bm S$ is taken arbitrarily}
		\label{figure3}
\end{figure}

 This lemma demonstrates that for any matrix $\bm S \in \mathbb{R}^{N \times d}$, the log ratio of the volumes, i.e., $ \log (\Vol_d(\bm \Phi \bm S)/\Vol_d(\bm S))$, concentrates around its expectation with a probabilistic concentration inequality (\ref{detconcent}). We can verify the result of this lemma via Monte-Carlo simulations, as shown in Figure \ref{figure3}. Given any arbitrary $\bm S$, $N = 10000$, $d=50$, and $M$ from 100 to 5000, 1000 times Monte-Carlo simulations for values of $ \log (\Vol_d(\bm \Phi \bm S)/\Vol_d(\bm S))$ in correspondence with different $M$ is demonstrated in Figure \ref{figure3}.
The figure shows that most of the values of $ \log (\Vol_d(\bm \Phi \bm S)/\Vol_d(\bm S))$ indeed concentrate around its expected value.

\subsubsection{Step 2.Covering Numbers}
As mentioned, without loss of generality, we only consider the so-called "Unit-Norm" Grassmann manifold, that is, the corresponding matrix with respect to each point on Grassmann manifold has unit-norm columns. In this step, several lemmas are given as follows.


\begin{figure}[htbp]
\centering
\includegraphics*[width=0.9\textwidth]{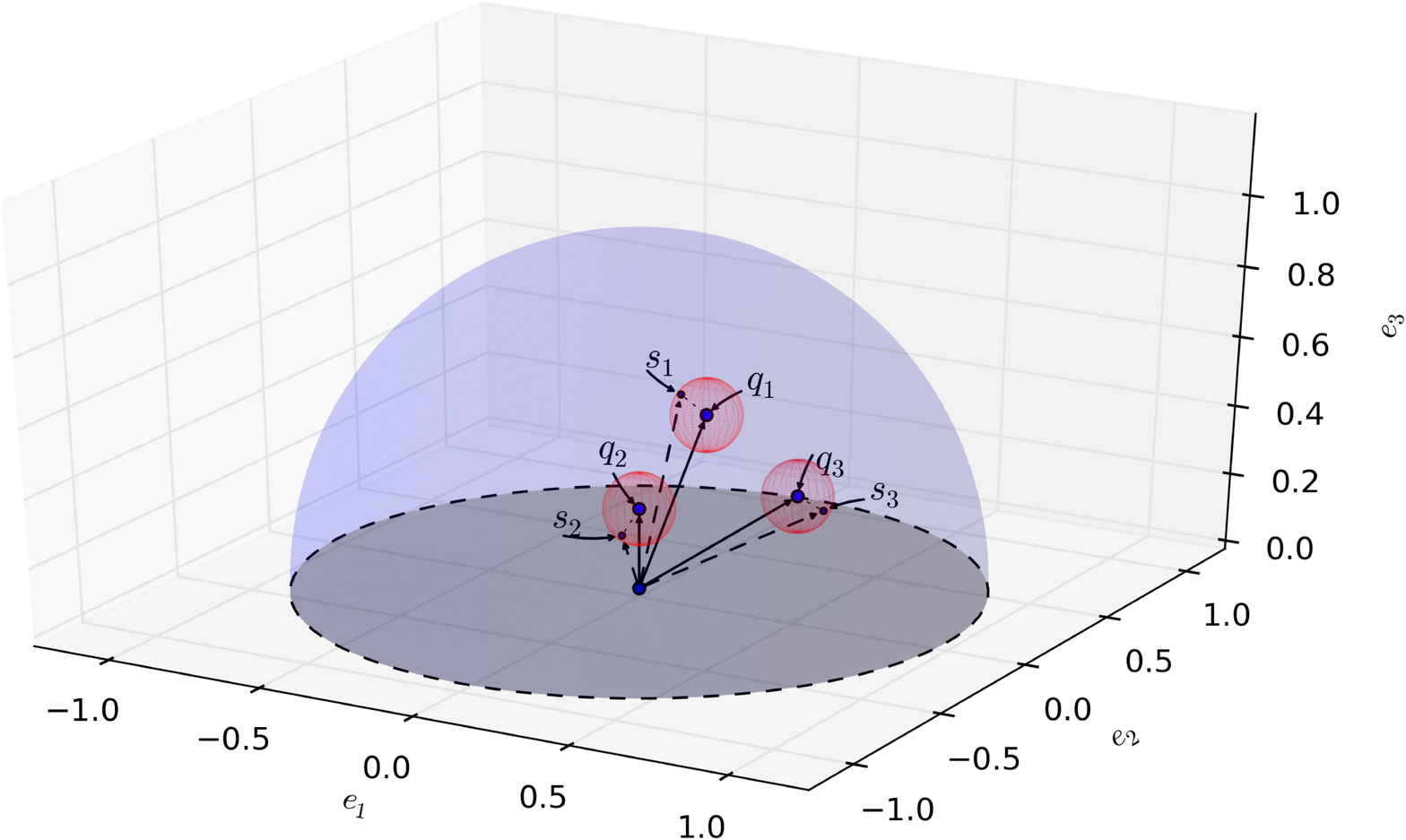}
\caption{Covering all of the unit norm Euclidean points $\bm s_1,\bm s_2, \bm s_3$ simultaneously with a finite number of balls centered at $\bm q_1,\bm q_2,\bm q_3$ in 3-dimensional Euclidean space}
		\label{figure4}
\end{figure}
\begin{Lemma}\label{Lemma2}
Given any point $\mathcal{X}$ on the "Unit-Norm" Grassmann manifold $\text{Gr}(k,N)$,  fix a constant $0<C_s<1$ and an integer $1\leq d \leq k$, there exists a constant $\delta_s^{(1)}>0$ depending on $C_s$. For any $0<\delta_0<\delta_s^{(1)}$, we have a finite set of matrices
\begin{equation}
\mbox{\boldmath{$\mathcal{Q}$}}=\{\bm{Q}_1,\cdots,\bm{Q}_m\} \nonumber
\end{equation}
where the cardinality $\#\mbox{\boldmath{$\mathcal{Q}$}}:=m$ only depends on $\delta_0$ and $d$, and $\bm Q_1, \cdots, \bm Q_m \in \mathbb{R}^{N \times d}$ are full-rank matrices with $\Span(\bm Q_1),\cdots,\Span(\bm Q_m)\subset \mathcal{X}$; such that for any matrix $\bm S\in\mathbb{R}^{N\times{d}}$ satisfying $\Span(\bm S) \subset\mathcal{X}$, $\Vol_d(\bm S)>C_s$, we can find a
$$\bm Q_r =[\bm q_1, \cdots \bm q_d] \in \mbox{\boldmath$\mathcal{Q}$},\ r=1,\cdots m ,$$ with $\bm q_j \neq \bm q_l,j \neq l$, and

\begin{equation}
\|\bm s_j-\bm q_j\|_2 \leq \delta_0,\quad j =  1 ,\cdots, d.
\end{equation}
The cardinality of \mbox{\boldmath$\mathcal{Q}$} satisfies $\#(\mbox{\boldmath$\mathcal{Q}$}) \leq \binom{\lfloor(3/\delta_0)^k\rfloor}{d}$.
\end{Lemma}
\text{Proof of Lemma \ref{Lemma2}: See Appendix \ref{App2}.}
\par This lemma states that, for all matrices $\bm S= [\bm s_1,\cdots, \bm s_d],  \Span(\bm S) \subset \mathcal{X}$ with unit-norm columns and $\Vol_d(\bm S)>C_s$, if a sufficiently small $\delta_0$ is chosen, we can always find a finite set $\mbox{\boldmath{$\mathcal{Q}$}}$ of matrices with different columns, such that each Euclidean point $\bm s_j$ on the unit sphere can be covered by at least one ball centered at $\bm q_j$ with radius $\delta_0$ ($1\leq j \leq d$).
Indeed, the theory of covering numbers states that for any given $\delta_0$, all unit-norm Euclidean points in a $k$-dimensional subspace $\mathcal{X}$ can be covered by a finite set of balls with radius $\delta_0$, and the cardinality of this finite set is bounded by$(3/\delta_0)^k$\cite{rudelson2010non}\cite{SimpleProof}. This lemma simultaneously covers different points $\bm s_1,\cdots, \bm s_d$ satisfying $\Vol_d(\bm S)>C_s$ with balls centered at different $\bm q_1,\cdots,\bm q_d$ for any given $0<C_s<1$. Obviously, the cardinality of $\mbox{\boldmath{$\mathcal{Q}$}}$ is bounded by the combination number of the cardinality $(3/\delta_0)^k$.
The intuition of Lemma \ref{Lemma2} is demonstrated in 3-dimensional Euclidean space in Figure \ref{figure4}.
\par Lemma \ref{Lemma2} shows that if the radius $\delta_0$ is notably small, $ \bm q_1,\cdots, \bm q_d$ will be highly close to $\bm s_1,\cdots, \bm s_d$. Thus, intuitively, we expect the volumes of $\bm Q_r$ and $\bm S$ to be arbitrarily close, which is stated in the following lemma.
\begin{figure}[htbp]
\centering
\includegraphics*[width=0.9\textwidth]{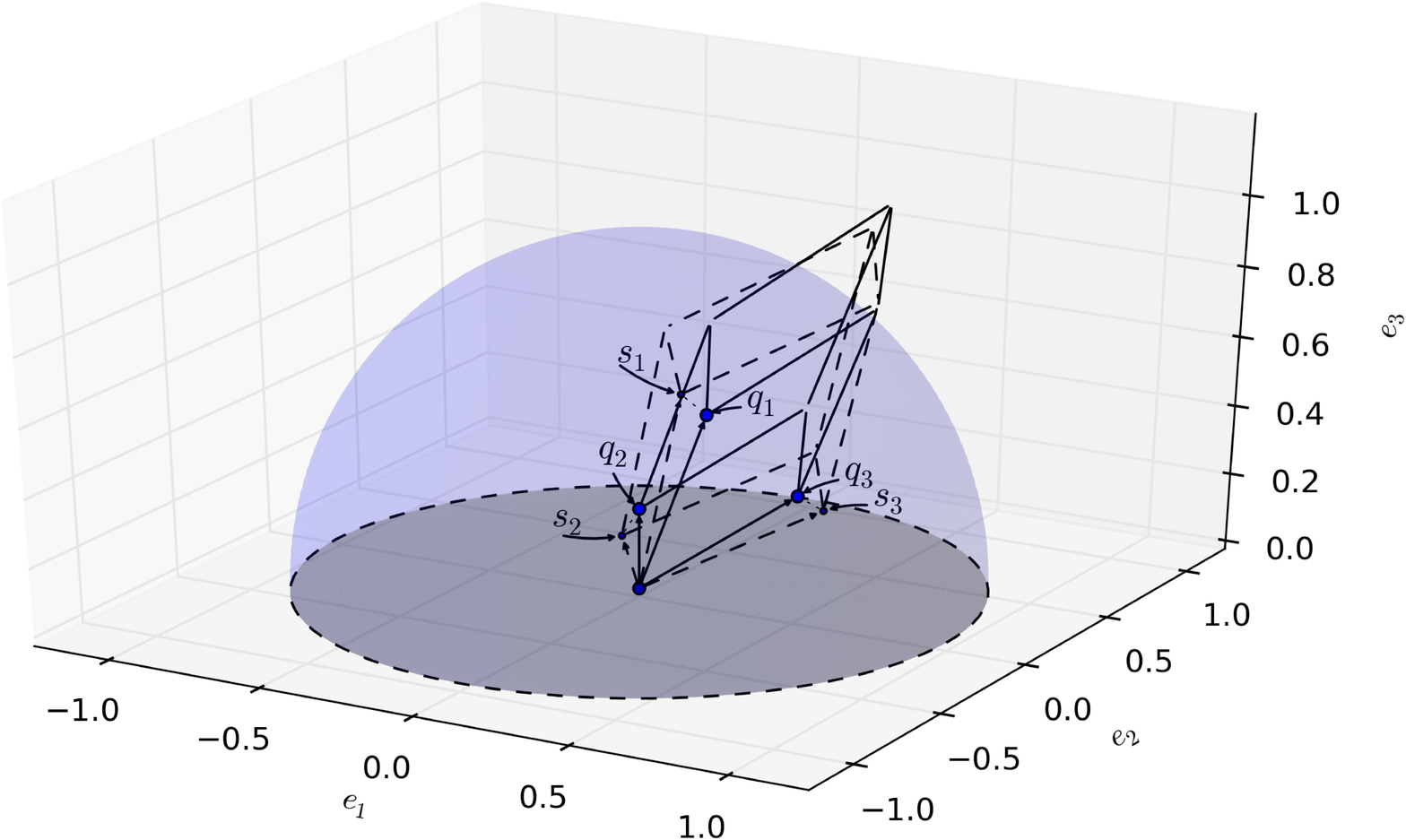}
\caption{The volume of a parallelotope spanned by $\bm q_1,\cdots,\bm q_d$ is similar to the volume spanned by $\bm s_1,\cdots, \bm s_d$}
		\label{figure5}
\end{figure}

\begin{Lemma}\label{Lemma3}
Given any point $\mathcal{X}$ on the "Unit-Norm" Grassmann manifold $\text{Gr}(k,N)$ and a random matrix $\bm \Phi \in \mathbb{R}^{M \times N}$ with elements $\phi_{i,j}$ being i.i.d Gaussian random variables with mean 0 and variance $1/M$,  fix a constant $0<C_s<1$ and an integer $1\leq d \leq k$, there exists a constant $\delta_s>0$ depending on $C_s$. For any $0<\delta_0<\delta_s$, we have a finite set of mattices
\begin{equation}
\mbox{\boldmath{$\mathcal{Q}$}}=\{\bm{Q}_1,\cdots,\bm{Q}_m\}, \nonumber
\end{equation}
where the cardinality $\#\mbox{\boldmath{$\mathcal{Q}$}}:=m$ only depends on $\delta_0$ and $d$, and $\bm Q_1, \cdots, \bm Q_m \in \mathbb{R}^{N \times d}$ are full-rank matrices with $\Span(\bm Q_1),\cdots,\Span(\bm Q_m)\subset \mathcal{X}$; such that for any matrix $\bm S\in\mathbb{R}^{N\times{d}}$ satisfying $\Span(\bm S) \subset\mathcal{X}$, $\Vol_d(\bm S)>C_s$, we can find a $\bm Q_r \in \mbox{\boldmath$\mathcal{Q}$}(r=1,\cdots,m)$, and

\begin{eqnarray}
\Vol_d(\bm Q_r)\cdot \exp(-d^{\frac{3}{2}}\delta_0/C_{1}) \leq& \Vol_d(\bm S) &\leq \Vol_d(\bm Q_r)\cdot \exp(d^{\frac{3}{2}}\delta_0/C_{2}), \label{VolRel1}\\
\Vol_d(\bm \Phi \bm Q_r)\cdot \exp(-d^{\frac{3}{2}}C_{\Phi}\delta_0/C_{1}) \leq& \Vol_d(\bm \Phi \bm S) &\leq \Vol_d(\bm \Phi \bm Q_r)\cdot \exp(d^{\frac{3}{2}}C_{\Phi}\delta_0/C_{2}). \label{VolRel2}
\end{eqnarray}
where $C_1,C_2>0$ are constant parameters related to $C_s$, and $0<C_{\bm\Phi}< \infty$ is a constant parameter related to matrix $\bm \Phi$.
In addition, the cardinality of \mbox{\boldmath$\mathcal{Q}$} satisfies $\#(\mbox{\boldmath$\mathcal{Q}$}) \leq \binom{\lfloor(3/\delta_0)^k\rfloor}{d}$.
\end{Lemma}
\par Lemma \ref{Lemma3} shows that because we can simultaneously cover all of the Euclidean points $\bm s_1,\cdots,\bm s_d$ that satisfy $\Vol_d(\bm S)>C_s$ with a finite set of balls centered at points $\bm q_1,\cdots,\bm q_d$ with radius $\delta_0$, then an arbitrarily small radius $\delta_0$ will ensure that $\Vol_d(\bm S)$ and $\Vol_d(\bm Q_r)$ are arbitrarily similar. The intuition of this lemma can be demonstrated in 3-dimensional Euclidean space in Figure \ref{figure5}.
\par According to these two lemmas, we can obtain the following lemma.

\begin{Lemma}\label{Lemma4}
Given any point $\mathcal{X}$ on the "Unit-Norm" Grassmann manifold $\text{Gr}(k,N)$ and a random matrix $\bm \Phi \in \mathbb{R}^{M \times N}$ with elements $\phi_{i,j}$ being i.i.d Gaussian random variables with mean 0 and variance $1/M$;  fix a constant $0<C_s<1$ and an integer $1\leq d \leq k$, there exists a constant $\delta_s>0$ depending on $C_s$. For any $0<\delta_0<\delta_s$, we have a finite set of matrices
\begin{equation}
\mbox{\boldmath{$\mathcal{Q}$}}=\{\bm{Q}_1,\cdots,\bm{Q}_m\}, \nonumber
\end{equation}
where the cardinality $\#\mbox{\boldmath{$\mathcal{Q}$}}:=m$ only depends on $\delta_0$ and $d$, and $\bm Q_1, \cdots, \bm Q_m \in \mathbb{R}^{N \times d}$ are full-rank matrices with $\Span(\bm Q_1),\cdots,\Span(\bm Q_m)\subset \mathcal{X}$; such that for any matrix $\bm S\in\mathbb{R}^{N\times{d}}$ satisfying $\Span(\bm S) \subset\mathcal{X}$, $\Vol_d(\bm S)>C_s$, we can find a $\bm Q_r \in \mbox{\boldmath$\mathcal{Q}$}(r=1,\cdots,m)$, and
\begin{equation}\label{CoverRes}
-d^{\frac{3}{2}} C' \delta_0 \leq \log \frac{\Vol_d(\bm \Phi \bm S)}{\Vol_d(\bm S)}-\log \frac{\Vol_d(\bm \Phi \bm Q_r)}{\Vol_d(\bm Q_r)} \leq d^{\frac{3}{2}} C' \delta_0,
\end{equation}
where $0<C'<\infty$ is a constant only depend on $C_s$ and $\bm\Phi$, and the cardinality of the set $\mbox{\boldmath$\mathcal{Q}$}$ satisfies
$$
\#\mbox{\boldmath{$\mathcal{Q}$}} \leq \binom{\lfloor(3/\delta_0)^k\rfloor}{d}.
$$
\end{Lemma}
\begin{IEEEproof}
According to Lemma \ref{Lemma2} and Lemma \ref{Lemma3}, we obtain
\begin{equation}
 \exp\{ -d^{\frac{3}{2}}(C_\Phi \delta_0/C_{1}+\delta_0/C_{2})\} \cdot \frac{\Vol_d(\bm \Phi \bm Q_r)}{\Vol_d(\bm Q_r)} \leq \frac{\Vol_d(\bm \Phi \bm S)}{\Vol_d(\bm S)} \leq \frac{\Vol_d(\bm \Phi \bm Q_r)}{\Vol_d(\bm Q_r)} \cdot \exp\{ d^{\frac{3}{2}}(C_\Phi \delta_0/C_{2}+\delta_0/C_1)\}.
\end{equation}

If we take $C' = \max\{C_\Phi /C_{1}+1/C_{2},C_\Phi/C_{2}+1/C_1\}$, then we obtain
\begin{equation}
 \exp\{ -d^{\frac{3}{2}}C'\cdot\delta_0\} \cdot \frac{\Vol_d(\bm \Phi \bm Q_r)}{\Vol_d(\bm Q_r)} \leq \frac{\Vol_d(\bm \Phi \bm S)}{\Vol_d(\bm S)} \leq \frac{\Vol_d(\bm \Phi \bm Q_r)}{\Vol_d(\bm Q_r)} \cdot \exp\{ d^{\frac{3}{2}}C'\cdot\delta_0\}.
\end{equation}
Lemma \ref{Lemma4} is now proved.
\end{IEEEproof}
\subsubsection{Step 3. Union Bound}
An immediate result from Lemma \ref{StepConcent} and Lemma \ref{Lemma4} is stated as follows:
\begin{Lemma}\label{Lemma5}
Consider any point on the Grassmann manifold $\mathcal{X} \in \text{Gr}(k,N)$, and a random matrix $\bm \Phi \in \mathbb{R}^{M \times N}$ with elements $\phi_{i,j}$ being i.i.d Gaussian random variables with mean 0 and variance $1/M$;
for any $0<C_s<1$ and any integer $1 \leq d \leq k$,  for every matrix $\bm S \in \mathbb{R}^{N \times d}, \Span(\bm S) \subset \mathcal{X}$, with unit-norm columns and $\Vol_d(\bm S)> C_s$, we state
that
 there exists $\delta_s>0$, and $C,C'>0$, only depend on $C_s$, such that for any $0 < \varepsilon<d^\frac{3}{2}\delta_s(1+C')$, we have:
\begin{equation}\label{VolConcentration}
\left| \log \frac{\Vol_d(\bm \Phi \bm S)}{\Vol_d(\bm S)}  -
\mathbb{E}\log \frac{\Vol_d(\bm \Phi \bm S)}{\Vol_d(\bm S)}\right| \leq \varepsilon,
\end{equation}
which holds with probability
\begin{equation}\label{VolConcentrationMeabd}
\mathbb{P} \geq 1 - 2\cdot \binom{\lfloor(3d^{\frac{3}{2}}(1+C')/\varepsilon)^k\rfloor}{d}\exp\Big\{-\varepsilon^2/\Big(4(1+C')^2\sum_{p=1}^d[\frac{1}{M-p+1}+C\frac{1}{(M-p+1)^2}]\Big)\Big\}.
\end{equation}
\end{Lemma}
\begin{IEEEproof}
 According to the results from Lemma \ref{Lemma4}, we know that for any given $0 < C_s < 1$ and any integer $1 \leq d \leq k$, for any given $0 < \delta_0 < \delta_s$ we can always find a finite set $\mbox{\boldmath{$\mathcal{Q}$}}$ such that (\ref{CoverRes}) holds for every matrix $\bm S$ , $\Span(\bm S) \subset \mathcal{X}$ with unit-norm columns and $\Vol_d(\bm S)> C_s$. Combining the result of Lemma \ref{StepConcent}, Lemma \ref{Lemma4} and the union bound, we obtain:

\begin{equation}
-\varepsilon' - d^{\frac{3}{2}}C'\delta_0 \leq \log \frac{\Vol_d(\bm \Phi \bm S)}{\Vol_d(\bm S)}  - \mathbb{E}\log \frac{\Vol_d(\bm \Phi \bm S)}{\Vol_d(\bm S)} \leq \varepsilon'+ d^{\frac{3}{2}}C' \delta_0
\end{equation}
holds for every matrix $\bm S$ and any $\varepsilon'>0$ with probability
\begin{equation}
\mathbb{P} \geq 1 - 2\cdot \binom{\lfloor(3/\delta_0)^k\rfloor}{d}\exp\Big\{-\varepsilon'^2/\Big(4\sum_{p=1}^d[\frac{1}{M-p+1}+C\frac{1}{(M-p+1)^2}]\Big)\Big\}.
\end{equation}
 If we take $\varepsilon' =d^{\frac{3}{2}} \delta_0$, and let $\varepsilon = (1+C' )d^{\frac{3}{2}}\delta_0$, then we produce (\ref{VolConcentration}) and (\ref{VolConcentrationMeabd}).
\end{IEEEproof}
\par Next, we finish the proof of Theorem \ref{MainThm}.\par
\textit{Proof of Theorem \ref{MainThm}:}
\par The result of Lemma \ref{Lemma5} shows the concentration inequality for all matrices in one point $\mathcal{X}_i$ on Grassmann manifold, and we can use the union bound to extend the result to every point from the set in Grassmann manifold $\boldsymbol{\mathcal{G}}(k,N,L)$. Thus, for every matrix $\bm S$ satisfying $\Span(\bm S) \subset \mathcal{X}_i$ in every point of the set $ \boldsymbol{\mathcal{G}}(k,N,L)$, with $\Vol_d(\bm S)> C_s$, (\ref{VolConcentration}) holds with probability
\begin{equation}
\mathbb{P} \geq 1 - 2L\cdot \binom{\lfloor(3d^{\frac{3}{2}}(1+C')/\varepsilon)^k\rfloor}{d}\exp\Big\{-\varepsilon^2/\Big(4(1+C')^2\sum_{p=1}^d[\frac{1}{M-p+1}+C\frac{1}{(M-p+1)^2}]\Big)\Big\}.
\end{equation}
\par Next, according to the Stirling's Inequality:
\begin{equation}
\binom{\lfloor(3d^{\frac{3}{2}}(1+C')/\varepsilon)^k\rfloor}{d} \leq (e\lfloor(3d^{\frac{3}{2}}(1+C')/\varepsilon)^k\rfloor/d)^d \leq (e\cdot d^{\frac{3}{2}k-1}\lceil(3(1+C')/\varepsilon)\rceil^k)^d,
\end{equation}
we state that if
\begin{eqnarray}
1/\Big(\displaystyle{\sum_{p=1}^d[\frac{1}{M-p+1}+C\frac{1}{(M-p+1)^2}]}\Big) \geq \hspace{0.4\textwidth} \nonumber \\
\frac{4(1+C')^2}{\varepsilon^2}\Big[ \log(2L) + d\cdot (\frac{3}{2}k - 1) \log (e d)+d\cdot k \log(\lceil \frac{3(1+C')}{\varepsilon}\rceil) +t \Big],\label{result1}
\end{eqnarray}
then $\mathbb{P} \geq 1 - e^{-t}$. Because

\begin{eqnarray}
\sum_{p=1}^d[\frac{1}{M-p+1}+C\frac{1}{(M-p+1)^2}] &\leq& \frac{d}{M-d+1}+\frac{C\cdot d}{(M-d+1)^2} \nonumber \\
&\leq & \frac{d}{M-d+1}(1+C).\label{result2}
\end{eqnarray}
Therefore, for the sufficient condition that (\ref{result1}) holds, we obtain
\begin{equation}
M \geq \frac{4(1+C')^2(1+C)\cdot d}{\varepsilon^2}\Big[\log(2L) + d\cdot(\frac{3}{2}k-1) \log(e \cdot d) + d\cdot k \log(\lceil\frac{3(1+C')}{\varepsilon}\rceil) +t\Big]+d-1.
\end{equation}
Thus, Theorem \ref{MainThm} is proved.
\subsection{Proof of Corollary \ref{Corr1}}

\par According to Theorem \ref{MainThm}, if we simultaneously consider two points $\mathcal{X}_i,\mathcal{X}_j \in \boldsymbol{\mathcal{G}}(k,N,L),1\leq i \neq j \leq L$ in the finite set $\boldsymbol{\mathcal{G}}(k,N,L)$ with $\mathcal{X}_i \bigcap \mathcal{X}_j = \{0\}$, then (\ref{expfor2k}) is a direct conclusion. Next, we know that
all of the $\bar{L} = L(L-1)/2$ linear subspaces $\mathcal{X}_i\oplus \mathcal{X}_j$ will form a new finite set in Grassmann manifold, i.e.,
$$\boldsymbol{\mathcal{G}}(2k,N,\bar{L}):=\{\mathcal{X}_i\oplus \mathcal{X}_j, 1\leq i \neq j \leq L\}.$$
 Next, for any given $0 < C_s <1$ and any dimension $0<d\leq 2k$, the Gaussian random measurement matrix $\bm \Phi \in \mathbb{R}^{M \times N}$ will provide the volume-based stable embedding for every matrix $\bm X \in \mathbb{R}^{N \times d}$, with $\Vol_{d}(\bm X)>C_s,\Span(\bm X) \subset \mathcal{X}_i \oplus \mathcal{X}_j$, which means that there exists $\delta_s>0$ and $C,C'>0$ such that for any
\begin{equation}\label{epsbd}
0 < \varepsilon<d^\frac{3}{2}\delta_s(1+C'),
\end{equation}
if
\begin{equation}\label{Mboundfor2k}
M \geq \frac{4(1+C')^2(1+C)\cdot d}{\varepsilon^2}\Big[ \log(2\bar{L}) + d\cdot(3k-1) \log(e \cdot d) + d\cdot 2k \log(\lceil\frac{3(1+C')}{\varepsilon}\rceil) +t\Big]+d-1,
\end{equation}
then
\begin{eqnarray}
-\varepsilon \leq \log \frac{\Vol_{d}(\bm \Phi \bm X)}{\Vol_{d}(\bm X)}  - \frac{1}{2} \sum_{p=1}^{d}\Big(\psi[(M-p+1)/2]+\log 2 - \log M\Big) \leq \varepsilon,
\end{eqnarray}
holds with probability
$\mathbb{P} \geq 1 - e^{-t}$.
\par Therefore, according to the union bound in probability, if we require the volume-based stable embedding property of all matrices $\bm X$ for all dimensions $1 \leq d \leq 2k$ and $\Vol_d(\bm X) > C_s$, the sufficient condition is that there exists $\delta_s>0$ and $C,C'>0$ such that for any $0 < \varepsilon<\delta_s(1+C')$ (i.e., less than the lowest bound in (\ref{epsbd}) when $d=1$), if $M$ satisfies the largest measurement bound for all $d$'s (i.e., the bound in (\ref{Mboundfor2k}) when $d=2k$), then the concentration inequality (\ref{conctr2k}) will hold with probability
\begin{equation}
\mathbb{P} \geq 1 - 2k\cdot e^{-t}.
\end{equation}
By replacing $t$ with $t+\log(2k)$, we obtain the result of Corollary \ref{Corr1}.
\subsection{Proof of Theorem \ref{MainPrinAgl}}
\par Theorem \ref{MainPrinAgl} is proven using the result of Corollary \ref{Corr1}. Consider every pair of points $\mathcal{X}_i$ and $\mathcal{X}_j$ in the set $\boldsymbol{\mathcal{G}}(k,N,L)$, if we take their unit norm basis $\bm X_i \in \mathbb{R}^{N \times k}$ and $\bm X_j \in \mathbb{R}^{N \times k}$,
satisfying $\Span(\bm X_i)= \mathcal{X}_i,\Span(\bm X_j)= \mathcal{\bm X}_j$ as well as $\Span([\bm X_i , \bm X_j])=\mathcal{X}_i \oplus \mathcal{X}_j$,
then for a given $0 < C_s < 1$, $\Vol_{2k}([\bm X_i , \bm X_j])>C_s$ for every $i \neq j$
\footnote{The existence of $C_s$ can be guaranteed by the disjointness of $\mathcal{X}_i$ and $\mathcal{X}_j$, which indicates $\Vol_{2k}([\bm X_i , \bm X_j]) \neq 0$.}.
The relationship between volume and principal angles implies
\begin{eqnarray}
\Vol_{2k}(\bm \Phi [\bm X_i,\bm X_j]) &=& \Vol_k(\bm \Phi \bm X_i)\cdot \Vol_k(\bm \Phi \bm X_j) \cdot \prod_m^k\sin\theta_m(\bm \Phi\mathcal{X}_i, \bm \Phi\mathcal{X}_j), \\
\Vol_{2k}([\bm X_i,\bm X_j]) &=& \Vol_k(\bm X_i)\cdot \Vol_k(\bm X_j) \cdot \prod_m^k\sin\theta_m(\mathcal{X}_i, \mathcal{X}_j).\label{volrel2}
\end{eqnarray}
Because of the unit-norm condition on the columns of $\bm X_i$ and $\bm X_j$, we have $\Vol_{k}(\bm X_i )\leq 1$ and $\Vol_{k}(\bm X_j ) \leq 1$, the relationship in (\ref{volrel2}) also indicates that $\Vol_{k}(\bm X_i ) > C_s $ and $\Vol_{k}(\bm X_j ) > C_s $,
and thus
\begin{equation}\label{anglerel}
\log \frac{\prod_m^k\sin\theta_m(\bm \Phi\mathcal{X}_i, \bm \Phi\mathcal{X}_j)}{\prod_m^k\sin\theta_m(\mathcal{X}_i, \mathcal{X}_j)} = \log \frac{\Vol_{2k}(\bm \Phi [\bm X_i,\bm X_j])}{\Vol_{2k}([\bm X_i,\bm X_j])} - \log \frac{\Vol_k(\bm \Phi \bm X_i)}{\Vol_k(\bm X_i)} - \log \frac{\Vol_k(\bm \Phi \bm X_j)}{\Vol_k(\bm X_j)} .
\end{equation}
Next, according to (\ref{conctr2k}) in Corollary \ref{Corr1}, if the measurement matrix $\bm \Phi$ provides volume-based stable embedding for every matrix $\bm X$ with every dimension $1\leq d \leq 2k$ and $\Vol_d(\bm X)>C_s$ in the set  $\boldsymbol{\mathcal{G}}(2k,N,\bar{L})$, then

\begin{eqnarray}
-\varepsilon \leq \log \frac{\Vol_{2k}(\bm \Phi [\bm X_i, \bm X_j])}{\Vol_{2k}([\bm X_i,\bm X_j])}  -  \frac{1}{2} \sum_{p=1}^{2k}\Big(\psi[(M-p+1)/2]+\log 2 - \log M\Big)\leq \varepsilon, \\
-\varepsilon \leq \log \frac{\Vol_{k}(\bm \Phi \bm X_i)}{\Vol_{k}(\bm X_i)}  -  \frac{1}{2} \sum_{p=1}^{k}\Big(\psi[(M-p+1)/2]+\log 2 - \log M\Big)\leq \varepsilon, \\
-\varepsilon \leq \log \frac{\Vol_{k}(\bm \Phi \bm X_i)}{\Vol_{k}(\bm X_j)}  -  \frac{1}{2} \sum_{p=1}^{k}\Big(\psi[(M-p+1)/2]+\log 2 - \log M\Big)\leq \varepsilon,
\end{eqnarray}
and combined with (\ref{anglerel}), we prove this theorem.
\section{Conclusion}
\par In this paper, by formulating subspaces as points on the Grassmann manifold, we studied the stable embedding of linear subspaces via Gaussian random matrices, and proposed a volume-preserving embedding property of measurement matrices based on the Grassmann manifold. The Grassmann manifold enables us to establish a new theoretical framework to study multi-dimensional signals.
In this paper, we proved a volume-based stable embedding of a finite set in Grassmann manifold via Gaussian random matrices. We showed that volumes of parallelotopes in every points of Grassmann manifold is preserved via Gaussian random measurement matrices. The number of compressive measurements required to ensure the stable embedding of Grassmann manifold with high probability was also obtained. This property is a multi-dimensional generalization of the conventional RIP or stable embedding property, which only concerns the preservation of length of vectors.
Additionally, we further explored the application of this volume-based stable embedding property to study the embedding effect on a generalized distance measure for compressed measurement signals on the Grassmann manifold.
We found that the generalized distance measure between compressed measurement signals on the Grassmann manifold, i.e., the product of principal sines, is well preserved via Gaussian random measurement matrices. Rigorous proof and discussions as well as numerical simulations were provided for validation.

\appendices
\section{Proof of Lemma \ref{StepConcent}}
\label{App1}
To prove Lemma \ref{StepConcent}, several preliminary results are required.
\begin{Lemma}\label{Lemma6}
Consider a Gaussian random matrix $\bm \Phi \in \mathbb{R}^{M \times N}, N > M $ with each entry $\phi_{i,j}$ satisfying $\phi_{i,j}\thicksim \mathcal{N}(0,1/M)$, For any full-rank matrix $\bm S = [\bm s_1,\bm s_2,\cdots,\bm s_d] \in \mathbb{R}^{N \times d}, d < M$, the volume of the parallelotope spanned by $\bm S \in \mathbb{R}^{N \times d}$ and $\bm \Phi \bm S \in \mathbb{R}^{M \times d}$ satisfies
\begin{equation}
\log\frac{\Vol_d (\bm \Phi \bm S)}{\Vol_d (\bm S)}\stackrel{F}{=}  \frac{1}{2}\log \det(\hat{\bm\Phi}_d^T \hat{\bm\Phi}_d),
\end{equation}
where $\bm{\hat \Phi_d} \in \mathbb{R}^{M \times d}$ is also a Gaussian random matrix with entries satisfying (\ref{MomentCond}), and the "F" above the equality means that the right side has the same distribution function as the left.
\end{Lemma}
\begin{IEEEproof}
\par From the condition of this Lemma, if the matrix $\bm S\in \mathbb{R}^{N \times d}$ has full column rank, then we can apply a singular value decomposition:
\begin{equation}
\bm S = \bm U \left[\begin{array}{c}
								\bm \Sigma_d \\
								\bm O
						\end{array} \right]\bm V^T,
\end{equation}
where $\bm U \in \mathbb{R}^{N \times N}, \bm V \in \mathbb{R}^{k \times k}$ are orthogonal matrices of the left and right singular vectors, and
$$\bm{ \Sigma}_d = \text{diag} (\sigma_1, \sigma_2, \cdots ,\sigma_d)$$
 is a diagonal matrix whose entries are singular values $\sigma_1, \sigma_2, \cdots ,\sigma_d$.
\par According to the definition of volume in (\ref{VolumeDef2}),
\begin{eqnarray}
\Big(\frac{\Vol_d (\bm \Phi \bm S)}{\Vol_d (\bm S)}\Big)^2 &=& \frac{\det (\bm S^T \bm \Phi^T\bm \Phi \bm S)}{\det (\bm S^T\bm S)} \nonumber \\
&=& \frac{\det(\bm V \left[ \bm \Sigma_d, \bm O \right] \bm U^T \bm \Phi^T \bm \Phi \bm U \left[\begin{array}{c}
								\bm \Sigma_d \\
								\bm O
						\end{array} \right]\bm V^T)}
						{\det(\bm V \left[ \bm \Sigma_d, \bm O \right] \bm U^T \bm U \left[\begin{array}{c}
								\bm \Sigma_d \\
								\bm O
						\end{array} \right]\bm V^T)}\nonumber \\
&=& \frac{\det(\bm V \bm \Sigma_d \left[ \bm I_d, \bm O \right] \bm U^T \bm \Phi^T \bm \Phi \bm U \left[\begin{array}{c}
								\bm I_d \\
								\bm O
						\end{array} \right]\bm \Sigma_d \bm V^T)}
						{\det(\bm V \bm \Sigma_d^2 \bm V^T)}\nonumber \\
&=& \frac{\det(\bm X_d^T \hat{\bm \Phi}_d^T\hat{\bm \Phi}_d \bm X_d)}
						{\det(\bm X_d^T \bm X_d)},	\label{JacobianVolume}					 
\end{eqnarray}
where
$$\hat{\bm \Phi}_d := \bm \Phi \bm U \left[\begin{array}{c}
								\bm I_d \\
								\bm O
						\end{array} \right] \in \mathbb{R}^{M \times d}, \ \bm X_d = \bm \Sigma_d \bm V^T \in \mathbb{R}^{d\times d}.$$
It is not difficult to prove that $\hat{\bm \Phi}_d \in \mathbb{R}^{N \times d}$ is still a Gaussian random matrix with entries satisfying (\ref{MomentCond}).						
\par Next with the knowledge of the multiplication property of the determinant of square matrices, we obtain
\begin{equation}
\sqrt{\frac{\det(\bm X_d^T \hat{\bm \Phi}_d^T\hat{\bm \Phi}_d \bm X_d)}
						{\det(\bm X_d^T \bm X_d)}} = \sqrt{\frac{\det(\bm X_d^T) \det(\hat{\bm \Phi}_d^T\hat{\bm \Phi}_d) \det(\bm X_d)}
						{\det(\bm X_d^T) \det(\bm X_d)}} = \sqrt{\det(\hat{\bm\Phi}_d^T \hat{\bm\Phi}_d)},
\end{equation}
and combined with (\ref{JacobianVolume}), the result of this lemma is proved.
\end{IEEEproof}
\begin{Lemma}\label{Lemma7}
(Bartelett Decomposition, \cite{TCai2013arXiv})
For a Gaussian random matrix $\hat{\bm \Phi}_d \in \mathbb{R}^{M \times d}, d < M$  with each entry $\phi_{i,j}$ satisfying $\phi_{i,j}\thicksim \mathcal{N}(0,1/M)$, the random variable $\log \det(\hat \Phi_d^T \hat \Phi_d)$ has the same distribution as the sum of $d$ independent $\log \chi^2$ random variables, that is:
\begin{equation}\label{sumlogchi}
\log \det(\hat \Phi_d^T \hat \Phi_d) \stackrel{F}{=} \sum_{p=1}^d\Big[\log(\mathcal{X}_{M-p+1}^2)-\log M \Big].
\end{equation}
The "F" above the equality indicates equality in distribution, and $\mathcal{X}_{M-p+1}^2$ denotes a chi-square random variable of order $M-p+1$.
\end{Lemma}
\par Combining the result of Lemma \ref{Lemma6} and Lemma \ref{Lemma7}, we prove Lemma \ref{StepConcent}.\par
\textit{Proof of Lemma \ref{StepConcent}:}
\par According to Lemma \ref{Lemma6} and Lemma \ref{Lemma7}, we must derive the concentration inequality of the sum of $d$ independent $\log \chi^2$ random variables in (\ref{sumlogchi}), because\cite{TCai2013arXiv}
\begin{equation}
\mathbb{E}\Big( \sum_{p=1}^d \log(\mathcal{X}_{M-p+1}^2) \Big) = \sum_{p=1}^d\Big[\psi[(M-p+1)/2]+\log 2\Big],
\end{equation}
where $\psi(x)$ is the Digamma function mentioned previously. Given that the entries of a Gaussian random matrix satisfy $\phi_{ij}\thicksim\mathcal{N}(0,1/M)$, we obtain
\begin{equation}
\mathbb{E}\{\log \frac{\Vol_d (\bm \Phi \bm S)}{\Vol_d (\bm S)}\} =  \frac{1}{2} \sum_{p=1}^{d}\Big(\psi[(M-p+1)/2]+\log 2 - \log M\Big).
\end{equation}
Thus, the problem becomes the concentration inequality for this random variable
\begin{equation}
Z := \log \frac{\Vol_d (\bm \Phi \bm S)}{\Vol_d (\bm S)} - \frac{1}{2} \sum_{p=1}^{d}\Big(\psi[(M-p+1)/2]+\log 2 - \log M\Big) \stackrel{F}{=}\sum_{p=1}^d \log(\mathcal{X}_{M-p+1}^2) -\sum_{p=1}^d\Big[\psi[(M-p+1)/2]+\log 2\Big].
\end{equation}
According to Markov's Inequality, we state
\begin{eqnarray}\label{Markov}
\mathbb{P}\{Z > \varepsilon\} = \mathbb{P}\{e^{\lambda Z} > e^{\lambda \varepsilon}\}\leq \frac{\mathbb{E}(e^{\lambda Z})}{e^{\lambda \varepsilon}},\quad \text{for any } \varepsilon > 0,\lambda >0,
\end{eqnarray}
where $\mathbb{E}(e^{\lambda Z}), \lambda \in \mathbb{R}$ is the Moment Generation Function. Thus (\cite{TCai2013arXiv}, A.7 of \cite{lee2012bayesian})
\begin{eqnarray}\label{MGF}
\lefteqn{\mathbb{E}(\exp(\lambda Z))}\\
&=& \prod_{p=1}^d \mathbb{E}(\exp(\lambda \log \chi^2_{M-p+1}))\cdot \frac{1}{\exp\{\lambda\big(\psi[(M-p+1)/2]+\log 2\big)\}} \nonumber \\
&=& \prod_{p=1}^d \mathbb{E}(\chi^2_{M-p+1})^\lambda\cdot \frac{1}{\exp\{\lambda\cdot \psi[(M-p+1)/2]\}\cdot 2^\lambda} \nonumber \\
&=& \prod_{p=1}^d \frac{\Gamma[(M-p+1)/2+\lambda]}{\Gamma[(M-p+1)/2]}\cdot 2^\lambda \cdot \frac{1}{\exp\{\lambda\cdot \psi[(M-p+1)/2]\}\cdot 2^\lambda} \nonumber \\
&=& \prod_{p=1}^d \frac{\Gamma[(M-p+1)/2+\lambda]}{\Gamma[(M-p+1)/2]} \cdot \frac{1}{\exp\{\lambda\cdot \psi[(M-p+1)/2]\}},
\end{eqnarray}
where $\Gamma(z)$ is the Gamma function. Taking the $\log$ of both sides, we obtain
\begin{equation}
\log \mathbb{E}\{\exp(\lambda Z)\} = \sum_{p=1}^d \Big(\log \Gamma[(M-p+1)/2+\lambda] -\log \Gamma[(M-p+1)/2] -\lambda \psi[(M-p+1)/2]\Big).
\end{equation}
If we use the asymptotic expansion of the Gamma function and Digamma function\cite{TCai2013arXiv}, we obtain
\begin{eqnarray}
\log \Gamma(z)&=& z\log z - z -\frac{1}{2} \log \frac{z}{2\pi} + \frac{1}{12z} + O(\frac{1}{|z|^2}) \\
\psi(z)&=& \log z - \frac{1}{2z}+O(\frac{1}{|z|^2}).
\end{eqnarray}
Using Taylor expansion, we obtain
\begin{eqnarray}\label{taylors}
\lefteqn{\log \Gamma[(M-p+1)/2+\lambda] - \log \Gamma[(M-p+1)/2] -\lambda \psi[(M-p+1)/2]} \nonumber \\
&=& \lambda \log[(M-p+1)/2]-\lambda \frac{1}{M-p+1}+\lambda^2 \frac{1}{M-p+1} \nonumber \\
 && -\lambda \log[(M-p+1)/2] + \lambda \frac{1}{M-p+1}+ O(\frac{\lambda^2}{(M-p+1)^2}) \nonumber \\
 &=& \lambda^2 \Big(\frac{1}{M-p+1} + O(\frac{1}{(M-p+1)^2})\Big).
\end{eqnarray}
Consider the remainder term
\begin{equation}
R_M := O(1/(M-p+1)^2)
\end{equation}
in (\ref{taylors}), for sufficiently large $M$, there exists $M_0 \in \mathbb{N}$, $C_0>0$ such that for all $ M > M_0$,
$$R_M \leq C_0/(M-p+1)^2.$$
If we take
$$C_M := R_M\cdot (M-p+1)^2, \ p \leq M \leq M_0$$
and let
\begin{equation}
C := \max\{C_p,\cdots, C_{M_0},C_0\},
\end{equation}
then
\begin{equation}
R_M \leq C/(M-p+1)^2
\end{equation}
holds for all $M\geq p$. Thus, the result in (\ref{MGF}) will become:
\begin{equation}
\mathbb{E}(\exp(\lambda Z)) \leq \exp\Big\{\lambda^2 \sum_{p=1}^d \Big[\frac{1}{M-p+1} + \frac{C}{(M-p+1)^2}\Big] \Big\}
\end{equation}
holds for a constant $C>0$, and (\ref{Markov}) becomes
\begin{equation}
\mathbb{P}\{Z > \varepsilon\} \leq \exp\Big\{-\lambda \varepsilon + \lambda^2 \sum_{p=1}^d \Big[\frac{1}{M-p+1} + \frac{C}{(M-p+1)^2}\Big] \Big\},
\end{equation}
which holds for any $\lambda > 0$. Thus, we can choose $\lambda$ such that
\begin{equation}
\mathbb{P}\{Z > \varepsilon\} \leq \arg\min_{\lambda >0}\bigg\{ \exp\Big\{-\lambda \varepsilon + \lambda^2 \sum_{p=1}^d \Big[\frac{1}{M-p+1} + \frac{C}{(M-p+1)^2}\Big] \Big\} \bigg\},
\end{equation}
If we take
\begin{equation}
\lambda_{\min} = \varepsilon/\bigg(2\cdot \sum_{p=1}^d \Big[\frac{1}{M-p+1} + \frac{C}{(M-p+1)^2}\Big]\bigg),
\end{equation}
then
\begin{equation}
\mathbb{P}\{Z > \varepsilon\} \leq  \exp\Big\{-\varepsilon^2/\big(\displaystyle{4 \sum_{p=1}^d \big[\frac{1}{M-p+1} + \frac{C}{(M-p+1)^2}\big]}\big) \Big\}.
\end{equation}
We can easily prove the same result for $\mathbb{P}\{-Z > \varepsilon\}$; as a result, Lemma \ref{StepConcent} is proved.
\section{Proof of Lemma \ref{Lemma2}}
\label{App2}
Lemma \ref{Lemma2} is a  direct derivation of the theory of covering numbers. From the knowledge of covering numbers \cite{SimpleProof}\cite{vershynin2010introduction}, for any given $\delta_0>0$ and any given $k$ dimensional linear subspace $\mathcal{X}$, there exists a set $\mathcal{Q}$ of finite elements with cardinality $\#(\mathcal{Q}) \leq \lfloor(3/\delta_0)^k\rfloor$, such that for every $\bm s \in \mathcal{X}, \|\bm s\|_2=1$, we can find at least one $\bm q \in \mathcal{Q}$ with $\|\bm q\|_2=1$ satisfying

\begin{equation}
\|\bm s-\bm q\|_2 \leq \delta_0.
\end{equation}
Then for any matrix $\bm S = [\bm s_1,\cdots,\bm s_d], \Span(\bm S) \subset \mathcal{X}$ with unit-norm columns and $ \Vol_d(\bm S)>C_s$, we can also find $\bm q_j$ for each $\bm s_j$, such that
\begin{equation}
\|\bm s_j-\bm q_j\|_2 \leq \delta_0,\quad j = 1,\cdots, d. \nonumber
\end{equation}
What we need to prove is that for an arbitrarily small $\delta_0$, these $\bm q_j$ in corresponding with $\bm s_j$ will be different for different $j$.
\par
As known from geometry, the volume of parallelotope spanned by $\bm S = [\bm s_1,\cdots,\bm s_d]$ equals the distance between any vector $\bm s_j$ and the hyperplane spanned by $\bm S_{\{l \neq j\}} := [\bm s_1,\cdots,\bm s_{j-1},\bm s_{j+1},\cdots,\bm s_d]$ multiplied by the volume of $\bm S_{\{l \neq j\}}$; that is:
\begin{eqnarray}
C_s^2 < \Vol_d^2(\bm S) &=& \det(\bm S^T \bm S) \nonumber \\
 &=& \det \Bigg(\left[\begin{array}{cc} \bm s_j^T \bm s_j & \bm s_j^T \bm S_{\{l \neq j\}} \\
													\bm S_{\{l \neq j\}}^T \bm s_j & \bm S_{\{l \neq j\}}^T \bm S_{\{l \neq j\}} \end{array}\right]\Bigg) \nonumber \\
&=& \det\big(\bm S_{\{k \neq j\}}^T \bm S_{\{l \neq j\}}\big)\cdot \det\Big(\bm s_j^T \bm s_j - \bm s_j^T \bm S_{\{l \neq j\}}\big(\bm S_{\{l \neq j\}}^T \bm S_{\{l \neq j\}}\big)^{-1}\bm S_{\{l \neq j\}}^T \bm s_j\Big) \nonumber \\
&=& \Vol_{d-1}^2\big(\bm S_{\{l \neq j\}}\big)\cdot \|\bm P_{\{l \neq j\}}^{\perp}\bm s_j\|_2^2,
\end{eqnarray}
where $\bm P_{\{l \neq j\}}^{\perp}:=\bm I_N - \bm S_{\{l \neq j\}}\big(\bm S_{\{l \neq j\}}^T \bm S_{\{l \neq j\}}\big)^{-1}\bm S_{\{l \neq j\}}^T$ is the matrix of projection onto the orthogonal completion of $\Span(\bm S_{\{l \neq j\}})$. Because of $\|\bm s_j\|_2=1, j = 1 ,\cdots, d$, using Hadamard's Inequality, we state $\Vol_{d-1}\big(\bm S_{\{l \neq j\}}\big)\leq 1$, and thus
\begin{equation}
\|\bm P_{\{l \neq j\}}^{\perp}\bm s_j\|_2^2 \geq \Vol_{d-1}^2\big(\bm S_{\{l \neq j\}}\big)\cdot \|\bm P_{\{l \neq j\}}^{\perp}\bm s_j\|_2^2 > C_s^2 .
\end{equation}
Intuitively, we also state
\begin{equation}\label{projectRel}
\|\bm P_{\{l \neq j\}}^{\perp}\bm s_j\|_2 \leq \|\bm P_l^{\perp}\bm s_j\|_2,\quad \forall l \neq j.
\end{equation}
The inequality (\ref{projectRel}) is not difficult to prove, because we know that
\begin{eqnarray}
\langle  \bm P_{l}\bm s_j,  \bm P_{\{l \neq j\}}^{\perp}\bm s_j \rangle = 0, \\
\langle  \bm P_{l}\bm s_j,  \bm P_{l}^{\perp}\bm s_j \rangle = 0,
\end{eqnarray}
so
\begin{equation}
\langle  \bm P_{l}\bm s_j,  \bm P_{l}^{\perp}\bm s_j - \bm P_{\{l \neq j\}}^{\perp}\bm s_j  \rangle = 0.
\end{equation}
and we obtain
\begin{equation}
\bm P_{l}\bm s_j + \bm P_{l}^{\perp}\bm s_j - \bm P_{\{l \neq j\}}^{\perp}\bm s_j = \bm s_j -\bm P_{\{l \neq j\}}^{\perp}\bm s_j = \bm P_{\{l \neq j\}}\bm s_j,
\end{equation}
thus
\begin{equation}
\|\bm P_{l}\bm s_j \|_2^2 + \|\bm P_{l}^{\perp}\bm s_j - \bm P_{\{l \neq j\}}^{\perp}\bm s_j\|_2^2 = \|\bm P_{\{l \neq j\}}\bm s_j\|_2^2.
\end{equation}
As a result $ \|\bm P_{\{l \neq j\}}\bm s_j\|_2^2 \geq \|\bm P_{l}\bm s_j \|_2^2$, because
$$\|\bm P_{\{l \neq j\}}\bm s_j\|_2^2 + \|\bm P_{\{l \neq j\}}^{\perp}\bm s_j\|_2^2=\|\bm P_{l}\bm s_j \|_2^2 + \|\bm P_{l}^{\perp}\bm s_j \|_2^2 = \|\bm s_j \|_2^2,$$
then (\ref{projectRel}) is proven, and we obtain
\begin{equation}
\|\bm P_{l}^{\perp} \bm s_j\|_2 \geq \|\bm P_{\{l \neq j\}}^{\perp}\bm s_j\|_2 > C_s,
\end{equation}
which holds for any $1 \leq j \neq l \leq d$. Because
\begin{equation}
\|\bm P_{l}^{\perp} \bm s_j\|_2^2 = \|\bm s_j\|_2^2 - |\langle \bm s_j, \bm s_l \rangle|^2 \|\bm s_l\|_2^2,
\end{equation}
If we let $\bm s_l := \bm s_j + \bm \delta_{j,l}$, with $\|\bm \delta_{j,l}\|_2:= \delta_{j,l}$, then
\begin{equation}
|\langle \bm s_j, \bm s_l \rangle|^2 = |\langle \bm s_j, \bm s_l \rangle + \langle \bm s_j, \bm \delta_{j,l} \rangle|^2 \geq (1-\delta_{j,l})^2,
\end{equation}
where the inequality comes from the Cauchy-Schwarz inequality $|\langle \bm s_j, \bm \delta_{j,l} \rangle| \leq \|\bm s_j\|_2 \|\bm \delta_{j,l}\|_2$ and $\|\bm s_j\|_2=1$, thus
\begin{equation}
1-(1-\delta_{j,l})^2 \geq \|\bm s_j\|_2^2 - |\langle \bm s_j, \bm s_l \rangle|^2 \|\bm s_l\|_2^2 > C_s^2,
\end{equation}
which means
\begin{equation}
1 + \sqrt{1-C_s^2}>\delta_{j,l} > 1-\sqrt{1-C_s^2}
\end{equation}
holds for any $1\leq j \neq l\leq d$; that is
\begin{equation}
\|\bm s_j - \bm s_l\|_2=\delta_{j,l} > 1-\sqrt{1-C_s^2},\quad \forall 1\leq j \neq l \leq d.
\end{equation}
\par Thus, we need only to take a certain $\delta_{s}^{(1)}$ that satisfies $\delta_{s}^{(1)} \leq (1-\sqrt{1-C_s^2})/2$, and for any $0<  \delta_0 \leq \delta_s^{(1)}$, we state
\begin{equation}\label{qdist}
\|\bm q_j - \bm  q_l\|_2 \geq \|\bm s_j-\bm s_l\|_2 - \|\bm s_j -\bm q_j\|_2 - \|\bm s_l - \bm q_l\|_2 > 1-\sqrt{1-C_s^2} - 2\delta_0 >0.
\end{equation}
In other words, if $\delta_0$ is sufficiently small, we can always find a group of different $\bm q_j$, such that the different $\bm s_j$ will be simultaneously covered by balls centered at different $\bm q_j$ with radius $\delta_0$.
From this point of view, the set $ \mbox{\boldmath$\mathcal{Q}$}$ is a subset that satisfies (\ref{qdist}) from all $d$ combinations of elements in $\mathcal{Q}$ with $\#(\mathcal{Q}) \leq \lfloor(3/\delta_0)^k\rfloor$. Thus, we obtain $\#(\mbox{\boldmath$\mathcal{Q}$}) \leq \binom{\lfloor(3/\delta_0)^k\rfloor}{d}$.
\section{Proof of Lemma \ref{Lemma3}}
Next, we prove Lemma \ref{Lemma3}. First, we consider (\ref{VolRel1}).
\par According to Lemma \ref{Lemma2}, for any $0 < C_s < 1$ and any integer $1 \leq d \leq k$, there exists $\delta_s^{(1)}>0$ such that for any $0 < \delta_0 \leq \delta_s^{(1)}$,
we can always find a finite set \mbox{\boldmath$\mathcal{Q}$} composed of matrices $\bm Q_r = [\bm q_1,\cdots,\bm q_d], \Span(\bm Q_r) \subset \mathcal{X}$, with  $\bm q_j \neq \bm q_l, 1\leq j \neq l \leq d$,
such that for all matrices $\bm S = [\bm s_1,\cdots,\bm s_d], \Span(\bm S) \subset \mathcal{X}$, with $\Vol_d(\bm S)>C_s$, there is a $\bm Q_r  \in$ \mbox{\boldmath$\mathcal{Q}$} that satisfies $\|\bm s_j-\bm q_j\|_2 \leq \delta_0,j = 1,\cdots, d$.
\par If we consider the matrix $\bm Q_r$ as a perturbation of  $\bm S$ by a matrix $\bm E$, where
\begin{equation}
\bm Q_r =\bm S + \bm E,
\end{equation}
and $\bm E = [\bm e_1,\cdots,\bm e_d],\|\bm e_j\|_2 \leq \delta_0,j = 1,\cdots, d$ is the perturbation matrix, then we can use matrix perturbation theory to analysis the relationship between the volumes of $\bm Q_r$ and $\bm S$.
\par
We denote $\sigma_1\geq \sigma_2 \geq \cdots \geq \sigma_d>0$ by the singular values of matrix $\bm S$, and $\tau_1\geq \tau_2 \geq \cdots \geq \tau_d>0$ by the singular values of matrix $\bm Q_r$, therefore, according to the Mirsky's Theorem of singular value perturbation (Theorem 4.11 of \cite{stewart1990matrix}), we obtain
\begin{equation}\label{perturbation}
|\sigma_l - \tau_l| \leq \|\bm S - \bm Q_r\|_2=\|\bm E\|_2,\quad l = 1,\cdots,d.
\end{equation}
\par According to the definition of matrix norm, we state
\begin{equation}
\|\bm E\|_2= \max_{\|\bm x\|_2=1}\{\frac{\|\bm E \bm x\|_2}{\|\bm x\|_2}\} = \sqrt{\lambda_{\max}(\bm E^T \bm E)},
\end{equation}
where $\lambda_{\max}(\bm E^T \bm E)$ is the maximum eigenvalue of matrix $\bm E^T \bm E$. Next, according to the theorem of Gershgorin's Circle\cite{horn2012matrix}, there is an integer $1\leq l \leq d$, such that
\begin{equation}
|\lambda_{\max}(\bm E^T \bm E)-\|\bm e_l\|_2^2|\leq  \sum_{j \neq l}^d |\bm e_l^T \bm e_j| \leq (d-1)\delta_0^2,
\end{equation}
so
\begin{equation}
\|\bm E\|_2 \leq \sqrt{d}\cdot \delta_0.
\end{equation}
Combined with (\ref{perturbation}), we obtain
\begin{eqnarray}
\tau_l - \sqrt{d}\delta_0 \leq \sigma_l \leq \tau_l + \sqrt{d}\delta_0, \label{perterb1}\\
\sigma_l - \sqrt{d}\delta_0 \leq \tau_l \leq \sigma_l + \sqrt{d}\delta_0\label{perterb2},
\end{eqnarray}
From the lemma's condition, we know that
\begin{equation}
\Vol_d(\bm S) = \prod_{l=1}^d \sigma_l  >C_s,
\end{equation}
and because
\begin{equation}
\sum_{l=1}^{d}\sigma_l^2 = tr(\bm S^T \bm S) = d.
\end{equation}
we obtain
\begin{equation}
\sum_{l=1}^{d-1}\sigma_l^2 = d - \sigma_d^2 \leq d,
\end{equation}
According to the inequality between the geometric average and arithmetic average,
\begin{eqnarray}
C_s^2 < \sigma_d^2 \cdot \prod_{l=1}^{d-1}\sigma_l^2  \leq \sigma_d^2 \cdot \Big( \frac{1}{d-1}\sum_{l=1}^{d-1}\sigma_l^2 \Big)^{\frac{1}{d-1}} \leq \sigma_d^2 \cdot  \Big( \frac{d}{d-1} \Big)^{\frac{1}{d-1}},
\end{eqnarray}
we obtain
\begin{equation}\label{Ssvalue}
\sigma_d \geq C_s \cdot  \Big( \frac{d}{d-1} \Big)^{-\frac{1}{2(d-1)}}.
\end{equation}
However, according to the left side of (\ref{perterb2}), we obtain
\begin{equation}
\tau_d \geq \sigma_d - \sqrt{d} \delta_0 \geq C_s \cdot  \Big( \frac{d}{d-1} \Big)^{-\frac{1}{2(d-1)}} - \sqrt{d} \delta_0.
\end{equation}
As a result, if we take a certain $\delta_s^{(2)} $ such that $0< \delta_s^{(2)} < \frac{C_s}{\sqrt{d}} \cdot  \Big( \frac{d}{d-1} \Big)^{-\frac{1}{2(d-1)}}$, then for any $\delta_0 \leq \delta_s:= \min\{\delta_s^{(1)},\delta_s^{(2)}\}$,
\begin{equation}\label{taudbound}
\tau_d \geq C_s \cdot  \Big( \frac{d}{d-1} \Big)^{-\frac{1}{2(d-1)}} - \sqrt{d} \delta_s^{(2)}.
\end{equation}
Then according to (\ref{perterb1}) and (\ref{perterb2}), we obtain
\begin{eqnarray}
\Vol_d(\bm S)=\prod_{l=1}^d \sigma_i &\leq& \prod_{l=1}^d(\tau_l + \sqrt{d}\cdot \delta_0) \nonumber \\
&=&\prod_{l=1}^d\tau_l \prod_{l=1}^d(1 + \sqrt{d}\cdot \delta_0/\tau_l),
\end{eqnarray}
According to (\ref{taudbound}), if we take
\begin{equation}
C_{2}:= C_s \cdot  \Big( \frac{d}{d-1} \Big)^{-\frac{1}{2(d-1)}} - \sqrt{d} \delta_s^{(2)},
\end{equation}
where $C_2$ is related to $C_s$, then $\tau_d > C_2$, which means
\begin{eqnarray}
\Vol_d(\bm S)&=&\Vol_d(\bm Q_r) \prod_{l=1}^d(1 + \sqrt{d}\cdot \delta_0/\tau_l) \nonumber \\
& < & \Vol_d(\bm Q_r) \prod_{l=1}^d(1 + \sqrt{d}\cdot \delta_0/C_{2}) \nonumber \\
& =& \Vol_d(\bm Q_r)\exp\{ \sum_{l=1}^d\log(1 + \sqrt{d}\cdot \delta_0/C_{2})\} \nonumber \\
& \leq & \Vol_d(\bm Q_r)\exp\{ d^{\frac{3}{2}}\cdot \delta_0/C_{2}\} .
\end{eqnarray}
The last inequality is due to the fact that $\log(1+x) \leq x$ for $x>0$. Thus, the right side of (\ref{VolRel1}) is proved. With knowledge of (\ref{perterb2}), we also state
\begin{eqnarray}
\Vol_d(\bm Q_r)=\prod_{l=1}^d \tau_l &\leq& \prod_{l=1}^d(\sigma_l + \sqrt{d}\cdot \delta_0) \nonumber \\
&=&\prod_{l=1}^d\sigma_l \prod_{l=1}^d(1 + \sqrt{d}\cdot \delta_0/\sigma_l),
\end{eqnarray}
and according to (\ref{Ssvalue}), if we take
\begin{equation}
C_1 := C_s \cdot  \Big( \frac{d}{d-1} \Big)^{-\frac{1}{2(d-1)}},
\end{equation}
then
\begin{eqnarray}
\Vol_d(\bm Q_r)&\leq &\prod_{l=1}^d\sigma_l \prod_{l=1}^d(1 + \sqrt{d}\cdot \delta_0/\sigma_l) \nonumber \\
&\leq &\Vol_d(\bm S)\prod_{l=1}^d(1 + \sqrt{d}\cdot \delta_0/C_1) \nonumber \\
&\leq & \Vol_d(\bm S)\exp\{ d^{\frac{3}{2}}\cdot \delta_0/C_{1}\} .
\end{eqnarray}
Thus, (\ref{VolRel1}) is now proved.
Next, we consider (\ref{VolRel2}). For a linear transform $\bm \Phi$, because all linear transforms are bounded linear operators, then there exists a constant $C_{\Phi}>0$, such that
\begin{equation}\label{bounded}
\|\bm \Phi \bm x\|_2 \leq C_{\Phi}\|\bm x\|_2,
\end{equation}
holds for all $\bm x \in \mathcal{X}$, where $\mathcal{X}$ is a given linear subspace.
It is noted that $\bm \Phi$ is an i.i.d. Gaussian random matrix with elements $\phi_{i,j}$ having zero mean and variance $1/M$, then (\ref{bounded}) holds almost surely for a sufficiently large $C_{\Phi}>0$, and $C_{\Phi}$ can be irrelevant to the dimension of $\bm \Phi$ \cite{rudelson2010non}. So we can generally state that $C_{\Phi}$ is a constant irrelevant to $M$ and $N$.
\par Next, if we denote $\hat \sigma_1\geq \hat \sigma_2 \geq \cdots \geq \hat \sigma_d>0$ by the singular values of matrix $\bm \Phi \bm S$ and $\hat \tau_1\geq \hat \tau_2 \geq \cdots \geq \hat \tau_d>0$ by the singular values of matrix $\bm \Phi \bm Q_r$, then similar to (\ref{perturbation}), we obtain
\begin{equation}
|\hat \sigma_l -\hat \tau_l| \leq \|\bm \Phi \bm S - \bm \Phi \bm Q_r\|_2=\|\bm \Phi \bm E\|_2.
\end{equation}
And we state $\|\bm \Phi \bm e_j\|_2 \leq C_{\bm \Phi}\delta_0,j = 1,\cdots, d$, thus, similarly we obtain
\begin{eqnarray}
\Vol_d(\bm \Phi \bm S)&\leq & \Vol_d(\bm \Phi \bm Q_r)\exp\{ d^{\frac{3}{2}}\cdot C_\Phi \delta_0/C_{2}\}, \\
\Vol_d(\bm \Phi \bm Q_r)&\leq &\Vol_d(\bm \Phi \bm S)\exp\{ d^{\frac{3}{2}}\cdot C_\Phi \delta_0/C_{1}\}.
\end{eqnarray}
Therefore, Lemma \ref{Lemma3} is now proved.

\bibliography{StableVolumeEmbedding_v2.bib}
\bibliographystyle {IEEEtran}
\end{document}